\documentclass[twocolumn,twocolappendix]{aastex63}

\graphicspath{{./}{figures/}}
\usepackage{color}
\usepackage{soul}
\usepackage{natbib}
\usepackage{hyperref}
\usepackage{amsmath}
\usepackage{xspace}
\usepackage{lineno}
\usepackage{graphicx}
\usepackage{enumitem}
\usepackage{amssymb}
\usepackage{xifthen}
\usepackage{hyperref}
\usepackage[normalem]{ulem}
\usepackage{comment}
\usepackage{float}
\usepackage{CJKutf8}
\usepackage{bbm}

\hypersetup{    
  colorlinks      = {true},
  linkcolor       = {blue},
  citecolor       = {blue},
  urlcolor        = {blue},
}

\newcommand{\code}[1]{\texttt{#1}\xspace}

\newcommand{\SSSSS}{${S}^5$\xspace}

\newcommand{\teff}{\ensuremath{T_\mathrm{eff}}\xspace}
\newcommand{\logg}{\ensuremath{\log\,g}\xspace}

\shorttitle{Mapping the Galactic Halo with BHBs}
\shortauthors{Yu, Li, Speagle et al.}

\begin{document}

\title{The Power of High Precision Broadband Photometry: Tracing the Milky Way Density Profile with Blue Horizontal Branch Stars in the Dark Energy Survey}
\author[0009-0008-3535-7356]{Fengqing Yu (\begin{CJK*}{UTF8}{gbsn}余枫青\ignorespacesafterend\end{CJK*})}
\affiliation{Department of Computer Science, University of Toronto, 40 St. George Street, Canada}
\affiliation{Data Sciences Institute, University of Toronto, 17th Floor, Ontario Power Building, 700 University Ave, Toronto, ON M5G 1Z5, Canada}

\author[0000-0002-9110-6163]{Ting S. Li}
\affiliation{David A. Dunlap Department of Astronomy \& Astrophysics, University of Toronto, 50 St George Street, Toronto ON M5S 3H4, Canada}
\affiliation{Dunlap Institute for Astronomy \& Astrophysics, University of Toronto, 50 St George Street, Toronto, ON M5S 3H4, Canada}
\affiliation{Data Sciences Institute, University of Toronto, 17th Floor, Ontario Power Building, 700 University Ave, Toronto, ON M5G 1Z5, Canada}

\author[0000-0003-2573-9832]{Joshua S. Speagle (\begin{CJK*}{UTF8}{gbsn}沈佳士\ignorespacesafterend\end{CJK*})}
\affiliation{Department of Statistical Sciences, University of Toronto, 9th Floor, Ontario Power Building, 700 University Ave, Toronto, ON M5G 1Z5, Canada}
\affiliation{David A. Dunlap Department of Astronomy \& Astrophysics, University of Toronto, 50 St George Street, Toronto ON M5S 3H4, Canada}
\affiliation{Dunlap Institute for Astronomy \& Astrophysics, University of Toronto, 50 St George Street, Toronto, ON M5S 3H4, Canada}
\affiliation{Data Sciences Institute, University of Toronto, 17th Floor, Ontario Power Building, 700 University Ave, Toronto, ON M5G 1Z5, Canada}

\author[0000-0003-0105-9576]{Gustavo E. Medina}
\affiliation{David A. Dunlap Department of Astronomy \& Astrophysics, University of Toronto, 50 St George Street, Toronto ON M5S 3H4, Canada}
\affiliation{Dunlap Institute for Astronomy \& Astrophysics, University of Toronto, 50 St George Street, Toronto, ON M5S 3H4, Canada}

\author[0000-0003-2644-135X]{Sergey~E.~Koposov}
\affiliation{Institute for Astronomy, University of Edinburgh, Royal Observatory, Blackford Hill, Edinburgh EH9 3HJ, UK}
\affiliation{Institute of Astronomy, University of Cambridge, Madingley Road, Cambridge CB3 0HA, UK}
\affiliation{Kavli Institute for Cosmology, University of Cambridge, Madingley Road, Cambridge CB3 0HA, UK}

\author[0000-0001-7516-4016]{Joss~Bland-Hawthorn}
\affiliation{Sydney Institute for Astronomy, School of Physics, A28, The University of Sydney, NSW 2006, Australia}
\affiliation{Centre of Excellence for All-Sky Astrophysics in Three Dimensions (ASTRO 3D), Australia}

\author[0000-0001-8536-0547]{Lara~R.~Cullinane}
\affiliation{Department of Physics and Astronomy, Johns Hopkins University, 3400 N. Charles St, Baltimore, MD 21218, USA}

\author[0000-0003-3734-8177]{Gwendolyn M. Eadie}
\affiliation{David A. Dunlap Department of Astronomy \& Astrophysics, University of Toronto, 50 St George Street, Toronto ON M5S 3H4, Canada}
\affiliation{Department of Statistical Sciences, University of Toronto, 9th Floor, Ontario Power Building, 700 University Ave, Toronto, ON M5G 1Z5, Canada}
\affiliation{Data Sciences Institute, University of Toronto, 17th Floor, Ontario Power Building, 700 University Ave, Toronto, ON M5G 1Z5, Canada}

\author[0000-0002-8448-5505]{Denis~Erkal}
\affiliation{Department of Physics, University of Surrey, Guildford GU2 7XH, UK}

\author[0000-0003-3081-93195]{Geraint~F.~Lewis}
\affiliation{Sydney Institute for Astronomy, School of Physics, A28, The University of Sydney, NSW 2006, Australia}

\author[0000-0002-9269-8288]{Guilherme~Limberg}
\affiliation{Universidade de S\~ao Paulo, IAG, Departamento de Astronomia, SP 05508-090, S\~ao Paulo, Brazil}

\author[0000-0003-1124-8477]{Daniel~B.~Zucker}
\affiliation{School of Mathematical and Physical Sciences, Macquarie University, Sydney, NSW 2109, Australia}
\affiliation{Macquarie University Research Centre for Astrophysics and Space Technologies, Sydney, NSW 2109, Australia}

\correspondingauthor{Ting S. Li}
\email{ting.li@astro.utoronto.ca}

\begin{abstract}
Blue Horizontal Branch (BHB) stars, excellent distant tracers for probing the Milky Way's halo density profile, are distinguished in the $(g-r)_0$ vs $(i-z)_0$ color space from another class of stars, blue straggler stars (BSs). 
We develop a Bayesian mixture model to classify BHB stars using high-precision photometry data from the Dark Energy Survey Data Release 2 (DES DR2). We select $\sim2100$ highly-probable BHBs based on their $griz$ photometry and the associated uncertainties, and use these stars to map the stellar halo over the Galactocentric radial range $20 \lesssim R \lesssim 70$ kpc. After excluding known stellar overdensities, we find that the number density $n_\star$ of BHBs can be represented by a power law density profile $n_\star \propto R^{-\alpha}$ with an index of $\alpha=4.28_{-0.12}^{+0.13}$, consistent with existing literature values. In addition, we examine the impact of systematic errors and the spatial inhomogeneity on the fitted density profile. 
Our work demonstrates the effectiveness of high-precision $griz$ photometry in selecting BHB stars. The upcoming photometric survey from the Rubin Observatory, expected to reach depths 2-3 magnitudes greater than DES during its 10-year mission, will enable us to investigate the density profile of the Milky Way's halo out to the virial radius, unravelling the complex processes of formation and evolution in our Galaxy.
\end{abstract}




\section{Introduction} \label{sec:intro}

The Milky Way halo contains fundamental information about the evolution history of our Galaxy and nature of dark matter \citep{helmi_stellar_2008}. It is generally believed that the Milky Way is formed via hierarchical formation, where small stellar systems such as dwarf galaxies are merged and assembled \citep{fukushima_stellar_2019, searle_composition_1978}. Hence, the stellar halo preserves a fossil record of the Galaxy's formation history and past accretion events \citep{helmi2020}. Moreover, the stellar halo can provide insight into the structure of the dark matter halo through stellar dynamics \citep{gerhard_dark_2012}. In upcoming surveys, it may even be possible to measure the Galaxy's change in mass with cosmic time by using the kinematics of a smoothly distributed halo population \citep{sharma23}, assuming such a population exists.

The Milky Way stellar halo can be directly probed with tracer populations such as red giant-branch (RGB) stars, RR Lyrae (RRL) stars, and blue horizontal-branch (BHB) stars. These stars are bright, and can be observed even in the very periphery of the Milky Way. Among them, $\mathrm{BHB}$ stars are frequently used because their absolute magnitudes (and thus distances) are relatively straightforward to calibrate \citep{Preston1991, fukushima_structure_2018}. The stellar density profile of the halo $n_\star(R)$ is often fitted with an inverse power law with index $\alpha$, such that $n_\star \propto R^{-\alpha}$, where $R$ is the Galactocentric radius. \citet{deason2011} mapped BHB using Sloan Digital Sky Survey Data Release 8 from 1 - 40 kpc with a broken
power-law model; they found an inner slope $\alpha_{\mathrm{in}} = 2.3$ and $\alpha_{\mathrm{in}} = 4.6$ with the break radius $\approx$ 27 kpc. 
\citet{2018ApJ...852..118D} subsequently mapped $\mathrm{BHB}$ stars starting at 50 kpc using Hyper Suprime-Cam (HSC) photometry; they measured a slope of 4 when excluding the Sagittarius (Sgr) stream, consistent with the power law from smaller distances.

On the other hand, \citet{thomas_-type_2018} showed an inner slope of $\alpha \approx 4.2$ and a shallower outer slope of $\alpha \approx 3.2$ beyond a radius of about 40 kpc by mapping $\mathrm{BHB}$ stars in the Canada-France Imaging Survey I. More recently, addressing the dichotomy in literature values of the single breaking radius, \citet{han_stellar_2022} showed a doubly broken power law with break radii at 12 kpc and 28 kpc using the H3 Survey \citep{2019ApJ...883..107C}. Because of the discrepancies in the slope and break radius, the stellar halo density profile continues to be an active area of research.

\begin{figure*}
    \centering
    \includegraphics[width=0.98\textwidth]{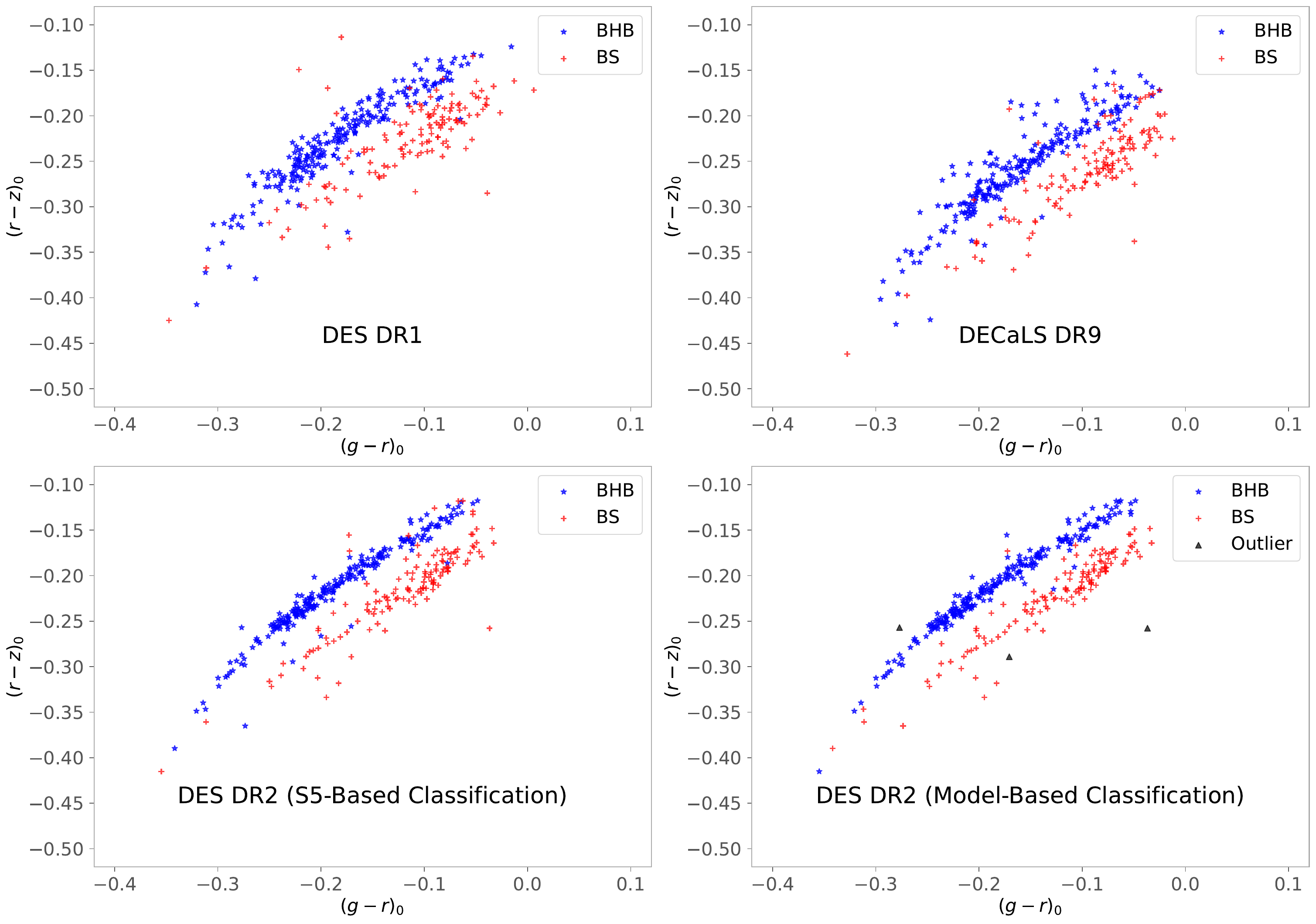}
    \caption{Photometry from DES DR1 (top left), DECaLS DR9 (top right), and DES DR2 (bottom left) cross-matched with \SSSSS survey targets, with target classes of BHB and BS as classified by \SSSSS surface gravity and temperature measurements (see Section \ref{sec:Data}). The bottom right panel uses the same photometry as the bottom left, but classified according to the photometric mixture model described in Section \ref{sec:model_classification}. The improved data quality of DES DR2 relative to DES DR1 and DECaLS DR9 is clearly visible, with both BHB and BS sequences becoming tighter and their separation more apparent. The high quality photometry in DES DR2 is one of the key factors that enables our mixture model to accurately recover the underlying BHB and BS populations from photometric colors alone. We note that we do not include $i-$band photometry here for comparison since DECaLS DR9 does not include $i-$band data.
    }
    \label{fig:photometry_comparison}
\end{figure*}

While BHBs are incredibly useful, finding them in photometric data can be challenging. To select BHBs from photometric data, the main obstacle lies in the removal of other contaminants that have similar colors and magnitudes to BHBs, such as blue straggler stars (BSs), white dwarfs (WDs), and quasi-stellar objects (QSOs). Among them, the removal of BSs remains the most difficult as these are much closer to BHBs in the color-color space than other contaminants. 
Previous works have largely relied on the $u-g$ color to distinguish BHBs and BSs \citep{yanny2000, deason2011}. \citet{2018ApJ...852..118D} used a combination of $g,r,i,z$ photometry from HSC to identify BHB stars.
Using the BHB and BS classifications based on surface gravity and effective temperature from the Southern Stellar Stream Spectroscopic Survey (\SSSSS) and photometry from the Dark Energy Survey Data Release 1 (DES DR1; \citealt{2018ApJS..239...18A}), \citet{2019MNRAS.490.3508L} identified a distinct and separable sequence for BHB and BS stars in  $(i-z)$ vs $(g-r)$ color-color space. 
However, this separation becomes more difficult for fainter stars because of the increase in uncertainties, and the decreased fraction of BHBs at fainter magnitudes.

In this study, we exploit the BHB/BS separation seen in \citet{2019MNRAS.490.3508L} and present a statistical model to classify BHB stars using photometric data from Dark Energy Survey Data Release 2 \citep[DES DR2;][]{2021ApJS..255...20A}, incorporating photometric uncertainties. Using BHB candidates selected from our model, we measure the Galactic stellar halo density profile $n_\star(R)$ and compare its slope with literature values.

In Section \ref{sec:Data}, we present an overview of the photometric data and preprocessing methods used in this study. In Section \ref{sec:model_classification}, we describe our statistical model that predicts the probability of a star being a BHB based on its colors and associated uncertainties. In Section \ref{sec:Density}, we use the selected $\mathrm{BHB}$ candidates to derive the Milky Way density profile. In Section \ref{sec:discuss}, we discuss our findings and implications, and we conclude in Section \ref{sec:conclusion}.

\section{Data} \label{sec:Data}

\subsection{High-quality Photometry}

High-precision photometry is required for teasing out the BHB signal from the contaminants for a large dataset. Here, we consider the photometry from the Dark Energy Camera \citep[DECam;][]{decam2015}, given its wide sky coverage, large telescope aperture, and the clear BHB/BS sequence separation identified by \citet{2019MNRAS.490.3508L}. We utilize two DECam-based surveys, the Dark Energy Survey (DES) and the DECam Legacy Survey (DECaLS), and compare the results from the public data releases of these two surveys. As part of the evaluation, we use the BHB/BS classification based on the measurements from \SSSSS. Throughout the paper, the subscript 0 indicates our use of extinction-corrected photometry; this correction is made possible using the dust map derived by \citet{Schlegel:1998}.

The DES is a wide-field optical/near-infrared imaging survey that contains 400 million astronomical objects \citep{2016MNRAS.460.1270D}, covering $\sim$5000 deg$^2$ of the South Galactic Cap region and obtaining photometry in $g,r,i,z,Y$ bands. In this comparison, we include both DES DR1 and DES DR2, the former of which is also used by \citet{2019MNRAS.490.3508L}.

We also exploit the DECaLS survey, which forms part of the DESI Legacy Imaging Surveys \citep{2019AJ....157..168D}. The latter mapped 14,000 deg$^2$ of the extragalactic sky visible from the northern hemisphere in three optical bands ($g, r$, and $ z$). DECaLS data release 9 (DECaLS DR9) incorporates the DES imaging, which is used in the comparison here.

\subsection{Spectroscopic Crossmatching}
\SSSSS \citep{2019MNRAS.490.3508L} is a spectroscopic survey with an initial focus on identifying stream member stars within the footprint of the DES, and an eventual goal of mapping the entire Southern sky. Its first public data release \citep[\SSSSS DR1;][]{S5DR1} includes a total of $\sim31000$ stars, most of which are concentrated on streams within the DES footprint. In \SSSSS DR1, the survey measures stellar parameters by fitting interpolated stellar atmosphere models to the IR spectra in the Calcium Triplet region. As shown in Figure 11 of \citet{2019MNRAS.490.3508L}, the stellar parameters of surface gravity and effective temperature from \SSSSS DR1 can effectively provide a distinction between BHB stars and BS stars, with BS stars having higher surface gravity at the same temperature.

We cross-match stars in \SSSSS DR1 with DES DR1, DES DR2, and DECaLS DR9 respectively. Following \citet{2019MNRAS.490.3508L}, We restrict our selection to stars with effective temperature $6000 \mathrm{K} < \teff < 10000 \mathrm{K}$ and surface gravity $2.5 < \logg < 6$. We then apply photometric cuts of $16 < g_0 < 19$, $-0.35 < (g-r)_0 < -0.05 $ and $-0.5 < (i-z)_0 < 0.1$. To ensure reliable stellar parameters, we require the stars in \SSSSS to have high signal-to-noise ratio (S/N) spectra ({\tt sn\_1700d} $>5$) and small uncertainties on surface gravity ({\tt logg\_std} $<0.5$). Furthermore, we require that all $g, r, i, z$ photometry are available in DES DR2.
RR Lyrae stars are removed by cross-matching with the RR Lyrae catalog from Gaia Data Release 3 \citep{2023A&A...674A...1G}. After the cut, a total of 365 stars remain. 
We adopt the same criterion defined by \citet{2019MNRAS.490.3508L} for BHB and BS classification in $\logg - \teff$ space.
In Figure ~\ref{fig:photometry_comparison}, we depict the stars in $(r-z)_0$ vs $(g-r)_0$ color-color space\footnote{Note the $i$-band is not used here because DECaLS DR9 does not contain $i$-band photometry}, with photometry from DES DR1, DECaLS DR9, and DES DR2 respectively.

Compared to both DES DR1 and DECaLS DR9, DES DR2 shows tighter sequences of BHB and BS stars and a clearer separation. The improvement in photometric calibration in DES DR2 enables our mixture model described in Section \ref{sec:model_classification} to differentiate the BHB and BS sequences {\it with photometry alone.} This is an important result. 

\begin{figure}
    \centering
    \includegraphics[width=0.48\textwidth]{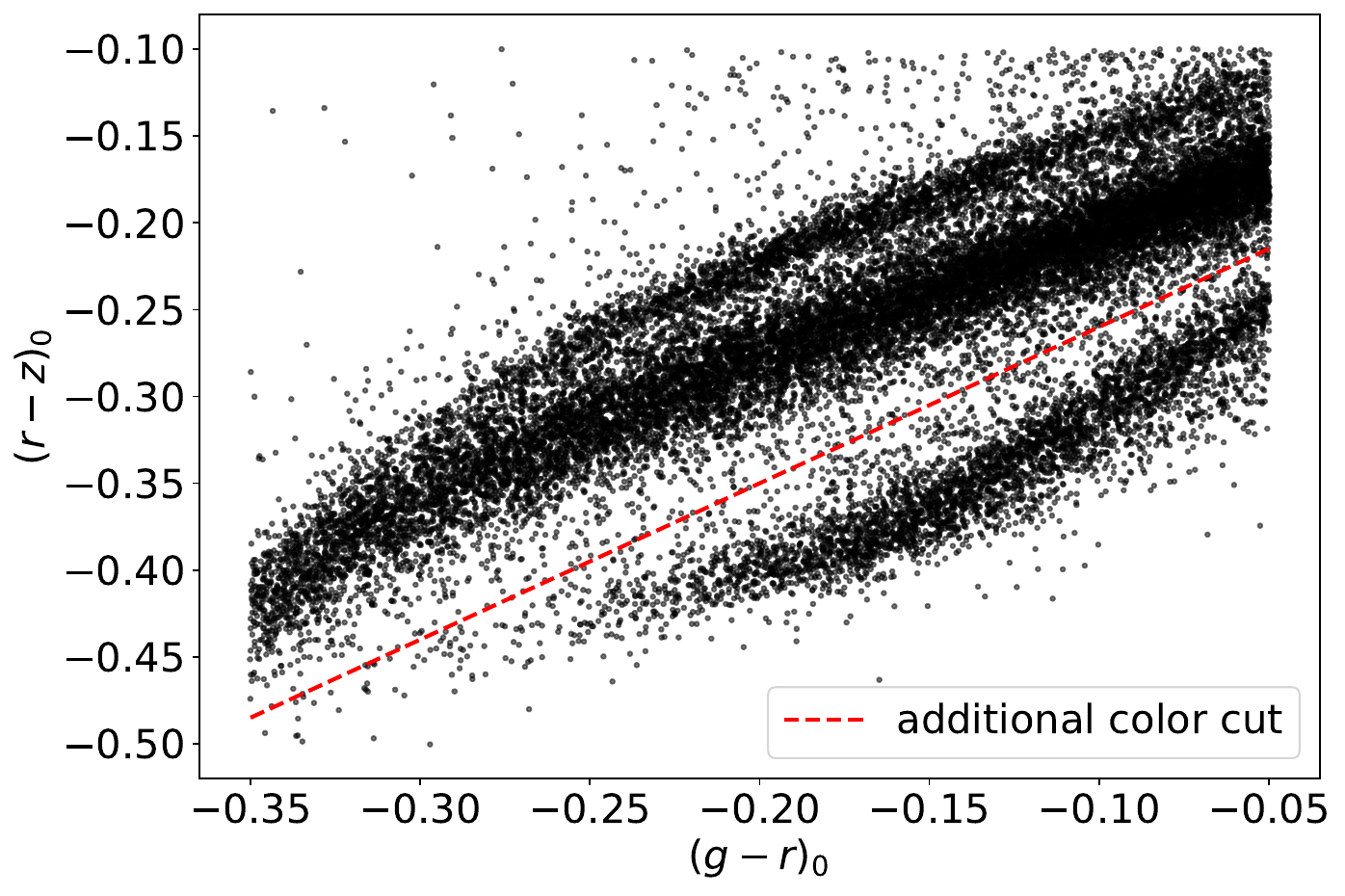}
    \caption{Extinction-corrected $(g-r)_0$ vs $(i-z)_0$ of DES DR2 photometry for stars with magnitudes in the range $18\leq g_0 \leq 20)$. The color-color cut used to remove contamination from WDs and quasars (QSOs) (lower sequence) is shown as a dashed red line.
    }
    \label{fig:data_cut}
\end{figure}

\begin{figure*}
    \centering
    \includegraphics[width=0.98\textwidth]{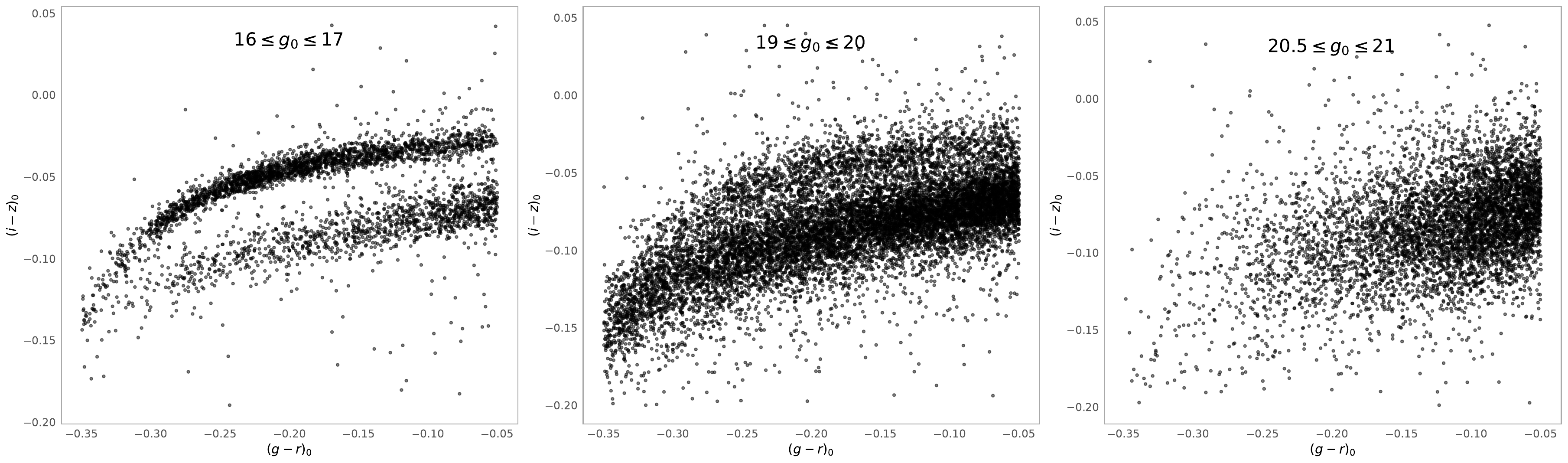}
    \caption{As Figure \ref{fig:photometry_comparison}, but now showing the distribution of stars in three different extinction-corrected $g_0$ magnitude ranges. As we extend into fainter magnitudes, the larger measurement errors cause the two well-defined sequences of BHB and BS stars to blend together, making it harder to disentangle potential BHB candidates. This motivates the use of the probabilistic mixture model described in Section \ref{sec:model_classification}.}
    \label{fig:BHB_magnitude}
\end{figure*}

\subsection{Data Preparation}\label{sec:dataprep}
Our ultimate goal is to develop a model which allows a purely photometric selection of BHB stars, since this allows us to select a much larger sample of these stars than those observed by spectroscopic surveys.
We build our model with DES DR2. We first apply a magnitude cut between 15 mag and 24 mag for $(g,r,i,z)$ photometry. Next, high-quality photometry is selected by using {\tt flags\_g,r,i,z}$< 4$. We use {\tt extended\_class\_coadd} $\leq 1$ to select only objects highly likely to be stars. To select blue stars, we apply color cuts of $-0.35 < (g-r)_0 < -0.05 $ (as motivated by \SSSSS DR1 x DES DR2 in Figure \ref{fig:photometry_comparison}) and $-0.2 < (i-z)_0 < 0.05$. One additional color cut, $(r-z)_0 \geq -0.17+(g-r)_0 \times 0.09$, is used to remove outliers likely to be WDs and QSOs, as shown in Figure \ref{fig:data_cut}.

The stars are then separated into different magnitude bins, based on $g_0$ band magnitude. 
We show stars at the selected magnitude ranges in Figure \ref{fig:BHB_magnitude}. 
From left to right, as the magnitude increases and stars become fainter, the BHB sequence and BS sequence merge together and a clear separation is no longer observed. This phenomenon is mostly due to the increase in photometric uncertainty at fainter magnitudes. While uncertainties in both $(i-z)_0$ and $(g-r)_0$ increase as the magnitude increases, uncertainties in $(i-z)_0$ are approximately 3 times larger than those in $(g-r)_0$ for stars with BHB-like colors. At $g_0 \approx 21$, the mean uncertainty for $(i-z)_0$ reaches 0.035, but the mean uncertainty for $(g-r)_0$ is only around 0.012. Thus, we enforce an additional magnitude cut of $16 < g_0 < 21$.
In total, we selected 46031 sources for our mixture model described in Section \ref{sec:model_classification}.

\section{Mixture Model}\label{sec:model_classification}
In this section, we describe a mixture model that predicts the probability of a star being a BHB, dependent on the $g_0$ magnitude, $(g-r)_0$ and $(i-z)_0$ color. 

\subsection{General Form of the Likelihood} \label{subsec:like}

We assume that our observed data can be classified into three categories: BHBs, BSs, and other contaminating sources (which we will refer to as ``outliers''). Each of these groups occupies fractions $f_{\rm BHB}$, $f_{\rm BS}$ and $f_{\rm out}$ in our dataset, respectively, subject to the constraint that $f_{\rm BHB} + f_{\rm BS} + f_{\rm out} = 1$. We assume that the BHB and BS stars lie along sequences in $(g-r)_0$ and $(i-z)_0$ color-color space centered around central ridgelines $\mu_{\rm BHB}$ and $\mu_{\rm BS}$, with some amount of intrinsic scatter $\sigma_{\rm BHB}$ and $\sigma_{\rm BS}$ respectively, while outliers are drawn from an alternate distribution which we take to be uniform between some lower and upper color bounds $c_{\rm min}$ and $c_{\rm max}$. Finally, we assume our observed colors and magnitudes have uncertainties $\sigma_{\rm obs}$ which are accurately estimated by the photometric pipelines. 

This gives us the following statistical model. First, we sample the object class $Y$ based on the fractions of objects in each group given some extra data values $D$ (e.g., magnitudes):
\begin{align}
    Y | D \sim {\rm Categorical}(f_{\rm BHB}, f_{\rm BS}, f_{\rm out}) 
\end{align}
Then, given the result, we assign one of three different  color-color distributions to the intrinsic colors $X$:
\begin{align}
    X | Y={\rm BHB}, D &\sim {\rm Normal}(\mu_{\rm BHB}, \sigma_{\rm BHB}) \\
    X | Y={\rm BS}, D &\sim {\rm Normal}(\mu_{\rm BS}, \sigma_{\rm BS}) \\
    X | Y={\rm Outlier}, D &\sim {\rm Uniform}(c_{\rm min}, c_{\rm max})
\end{align}
For example, if the star is categorized as a BS, then the data is assumed to be distributed as normal with mean $\mu_\mathrm{BS}$ and standard deviation $\sigma_\mathrm{BS}$.

Finally, we add in observational uncertainties to get the distribution of the observed colors $X_{\rm obs}$:
\begin{equation}
    X_{\rm obs} | X \sim {\rm Normal}(X, \sigma_{\rm obs})
\end{equation}

Given a collection of $n$ independent and identically distributed (iid) observations $\lbrace X_{{\rm obs}, i} \rbrace_{i=1}^{n}$ and combining all our parameters that characterize the detailed relationships in our model into the vector $\boldsymbol{\theta}$, this gives us the total (combined) log-likelihood:
\begin{equation}
    \ln P_{\rm tot}(\boldsymbol{\theta}) = \sum_{i=1}^{n} \ln P(X_{{\rm obs}, i} | X_i, Y_i, D_i, \boldsymbol{\theta}).
\end{equation}
The individual likelihood for each object can be written as a sum of the class-dependent log-likelihoods over $j=\lbrace {\rm BHB}, {\rm BS}, {\rm Outlier} \rbrace$:
\begin{multline}
    P(X_{{\rm obs}, i} | X_i, Y_i, D_i, \boldsymbol{\theta}) = \\
    \sum_j \mathbbm{1}({Y_i = j}) P_j(X_{{\rm obs}, i} | X_i, D_i, \boldsymbol{\theta})
\end{multline}
where $\mathbbm{1}({Y_i = j})$ is the indicator function which gives $1$ when $Y_i = j$ and $0$ otherwise. $P_j(\cdot)$ is the probability distribution for $X$ within each class (either Normal or Uniform). 

Of course, in practice the particular class $Y_i$ for object $i$ is unknown. Therefore, we marginalize over all possible classes, which allows us to replace our indicator function with the expected fraction of objects in each class:
\begin{multline}
    P(X_{{\rm obs}, i} | X_i, D_i, \boldsymbol{\theta}) = \\ \sum_j f_j(D_i) \, P_j(X_{{\rm obs}, i} | X_i, D_i, \boldsymbol{\theta})
\end{multline}
In addition, the true color $X_i$ of a given object is not known. 
Thus, We must also marginalize over all possible colors given our observations. Since it is not guaranteed to obtain an analytic solution, we will write out the full likelihood below:
\begin{multline}
    P(X_{{\rm obs}, i} | D_i, \boldsymbol{\theta}) = \\ \sum_j f_j(D_i) \int P(X_{{\rm obs}, i} | X_i, D_i, \boldsymbol{\theta}) P_j(X_i | D_i, \boldsymbol{\theta}) \, {\rm d}X_i \label{eq:marginalize}
\end{multline}
This 
gives us a tractable form of the likelihood to evaluate, where: 
\begin{itemize}
    \item $P(X_{{\rm obs}, i} | X_i, D_i, \boldsymbol{\theta})$ is the noise distribution of the observed colors given the true colors, which we assume follow a Normal distribution with unknown mean and known measurement uncertainties $\sigma_{\rm obs}$.
    \item $P_j(X_i | D_i, \boldsymbol{\theta})$ describes the intrinsic color-color relationships for BHB and BS stars as well as outliers, which we take to also be Normal distributions centered around some mean ridgeline $\mu_j$ with some intrinsic scatter $\sigma_{\rm int}$ (for BHB and BS classes) or Uniform within some color boundaries (for outliers).
    \item The integral over all possible intrinsic colors $X_i$ is the explicit way in which we marginalize over the unknown true colors for a given object.
\end{itemize}

\subsection{Detailed Model Implementation}

To determine the final functional form for our model, we fit individual models over objects from distinct magnitudes bins over $g_0 = (16,21)$ with a width of 0.5 mag, as well as across the entire $g_0$ magnitude range. After exploring a number of different models with varying complexity, we find a fixed third-order polynomial can accurately describe the ridgeline for the BHB and BS populations at all magnitudes. We take this to be a model for $(i-z)_0$ color as a function of $(g-r)_0$ color, giving us four parameters each:
\begin{align} \label{eq:bhb}
    \mu_{\rm BHB} \equiv (i-z)_{0, {\rm BHB}} &= \sum_{k=0}^{3} a_k [(g-r)_0]^k \\
    \mu_{\rm BS} \equiv (i-z)_{0, {\rm BS}} &= \sum_{k=0}^{3} b_k [(g-r)_0]^k \label{eq:bs}
\end{align}
We also confirm that the data are fully consistent with constant intrinsic scatter terms $\sigma_{\rm BHB}$ and $\sigma_{\rm BS}$ across all magnitudes, giving us two additional parameters. We find the probability of having a source belong to each group can be described using a linear function of $g_0$ magnitude, giving us four additional parameters:
\begin{align}
    f_{\rm out}(g_0) &= c_0 + c_1 \times g_0 \\
    f_{\rm BHB}(g_0) &= (1 - f_{\rm out}(g_0)) \times (d_0 + d_1 \times g_0) \label{eq: BHB ratio}
\end{align}
with the probability of being in the BS class determined via:
\begin{equation}
    f_{\rm BS}(g_0) = 1 - f_{\rm BHB}(g_0) - f_{\rm out}(g_0)
\end{equation}
This gives us a total of 14 unique parameters in our model to parametrize the distribution of BHB, BS, and outlier sources.

To marginalize over the observational uncertainties in color, we solve the integral outlined in equation \eqref{eq:marginalize} in two parts. First, we note that the distribution of $(i-z)_0$ color at a given $(g-r)_0$ color is either Normal (BS or BHB) or uniform (outlier). The integral of a product of two normal distributions is analytic and gives a new normal distribution with a standard deviation of $\sqrt{\sigma_{(i-z)_0}^2 + \sigma_{\rm BHB}^2}$. This leaves us with just the integral over $(g-r)_0$. For each object, we define a grid of 50 points in $(g-r)_0$ evenly spaced between $\pm 5 \sigma_{(g_r)_0}$, compute the integral over $(i-z)_0$ given each $(g-r)_0$, and then compute the remaining 1-D integral using a simple Riemann sum over the associated $(g-r)_0$ likelihood. For the uniform case, the integral always gives a constant value that does not depend on the input parameters (and therefore can simply be ignored).

\begin{table*}
\caption{
Uniform prior range, best-sampled results with the maximum likelihood, and uncertainties of the 14 parameters in the photometric mixture model.
}
\label{table:Mixture Model Prior}
\centering
\begin{tabular}{c c c c }
\hline
Parameter  & Uniform Prior Range & Best Fit\footnote{Samples with the maximum likelihood}
 & 95$\%$ Credible Interval\\
\hline
\hline
        $a_3$ & $(-0.2,0.2)$ & $-$0.0088 & $(-0.0019,0.0020)$ \\
        $a_2$ & $(-1.0,1.0)$ & 0.438 & $(-0.037,0.037)$ \\ 
        $a_1$ & $(-10,10)$ & 2.68 & $(-0.21,0.21)$\\
        $a_0$ & $(-25,25)$ & 6.88 &       $(-0.36,0.36)$\\ 
        $\sigma_{\mathrm{BHB}}$ & $(0,0.015)$ &  0.0049  & $(-0.0001, 0.0001)$\\ 
        $b_3$ & $(-0.2,0.2)$ & $-$0.0512 & $(-0.0018, 0.0018)$\\
        $b_2$ & $(-1.0,1.0)$ & 0.357 & $(-0.037, 0.037)$\\
        $b_1$ & $(-10,10)$ & 1.54 & $(-0.22, 0.22)$\\ 
        $b_0$ & $(-25,25)$ & 3.74 & $(-0.40, 0.39)$\\
        $\sigma_{\mathrm{BS}}$ & $(0,0.015)$ &  0.0107 & $(-0.0002, 0.0002)$\\
        \hline
        $c_1$ & $(-3.5,3.5)$ & 0.149 & $(-0.015, 0.017)$\\
        $c_0$ & $(-0.6,0.6)$ & -0.0071 & $(-0.0008, 0.0007)$\\
        \hline
        $d_1$ & $(-3.5,3.5)$ &  3.022 & $(-0.070, 0.072)$\\
        $d_0$ & $(-0.6,0.6)$ & -0.1450 & $(-0.0037, 0.0036)$\\ 
\hline
\end{tabular}
\end{table*}

We assume broad, uniform priors on all parameters as listed in Table \ref{table:Mixture Model Prior}. We sample from posterior using the \texttt{dynesty} Nested Sampling package \citep{2020MNRAS.493.3132S,koposov_joshspeagledynesty_2023} version 1.2.3 in Dynamic Nested Sampling mode \citep{2004AIPC..735..395S,10.1214/06-BA127,higson_dynamic_2019} under default settings that used multi-ellipsoidal bounds and uniform sampling \citep{2009MNRAS.398.1601F}.

\subsection{Results}

\begin{figure*}[!htb]
    \centering
    \includegraphics[width=0.98\textwidth]{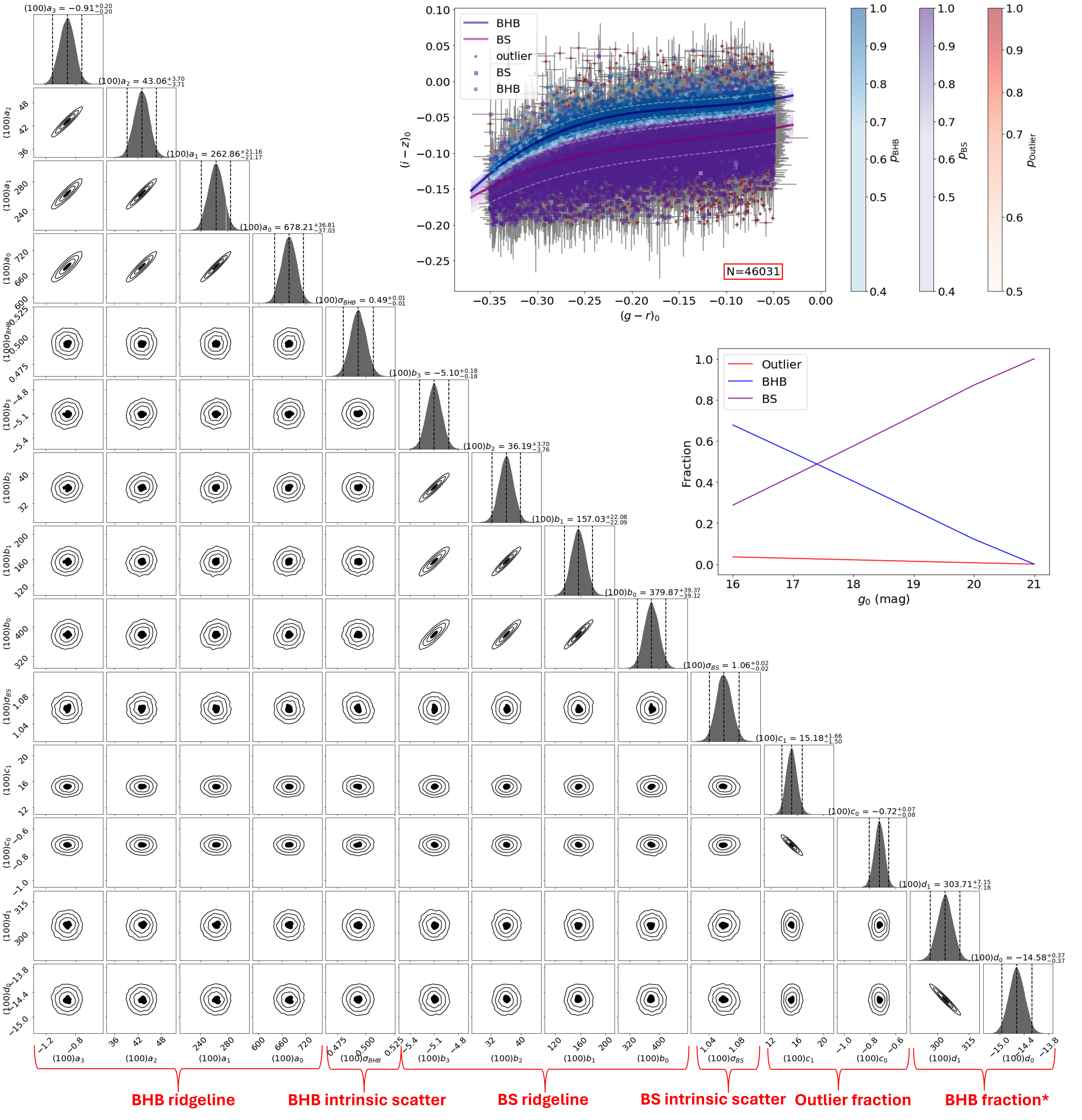}
    \caption{Bottom left: A corner plot depicting the 2D and 1-D marginal posterior probability distributions of the 14 parameters of our photometric mixture model estimated with \texttt{dynesty}. \textit{All values were multiplied by a factor of 100 for visualization purposes.} The parameters model different aspects of the data, as noted in red at the bottom. The parameter \textit{BHB fraction*} describes the ratio of BHBs to non-outliers. Top right: Stars classified as BHBs, BSs, and outliers are shown in blue circles, purple squares, and red stars, respectively. The predicted class is defined to be the one that shows the highest posterior-marginalized probability, and the color is used to represent the probability (one of $p_{\mathrm{BHB}}, p_{\mathrm{BS}}, p_{\mathrm{outlier}})$. The blue solid line and purple solid line show the ridgelines $\mu_{\mathrm{BHB}}, \mu_{\mathrm{BS}}$, calculated from the best-fit parameters in Table \ref{table:Mixture Model Prior}. The shaded regions with dashed lines on their boundaries along each ridgeline show the intrinsic scatters $\sigma_{\rm int}$ for each sequence.
    Middle right: The predicted ratio of BHB (blue), BS (purple), and outliers (red) as a function of magnitude. Overall, our model has parameters that are well-constrained, accurately trace the photometric distribution of stars, successfully identify photometric outliers, and generate physically-sensible magnitude dependencies across each subgroup. }
    \label{fig:Mixture_Model}
\end{figure*}

The result of the sampling process is shown in the bottom left panel of Figure \ref{fig:Mixture_Model}. The corner plot shows the 1D and 2D marginal posterior distributions of the 14 parameters in our photometric mixture model. There is some covariance among $a_3, a_2, a_1, a_0$. As those parameters collectively describe the position of the BHB ridgeline, they are expected to be dependent on each other. This holds true for parameters $b_3, b_2, b_1, b_0$ as well. In addition, the coefficients of the polynomial that models the outlier ratio should also be dependent on each other, corresponding to the observation of covariance in $c_1, c_0$. This applies to the pair $d_1, d_0$ as well. Other than those pairs, the rest of the parameters should not be dependent on each other, and indeed there is no evidence of covariance shown in the plot. 

Using the best-fit parameters in Table \ref{table:Mixture Model Prior}, the probability of each star being a BHB is calculated via
\begin{equation}
    p_{\mathrm{BHB}} = \frac{f_{\mathrm{BHB}} L_{\mathrm{BHB}}}{f_{\mathrm{BHB}} L_{\mathrm{BHB}}+ f_{\mathrm{BS}} L_{\mathrm{BS}}+f_{\mathrm{outlier}} L_{\mathrm{outlier}}}
\end{equation}
where $L_j$ is the corresponding likelihood described in Section \ref{subsec:like}. $p_{\mathrm{BS}}$ and $ p_{\mathrm{outlier}}$ are defined similarly.

In the top right of Figure \ref{fig:Mixture_Model}, we show a visualization of the predicted class probabilities of the stars used in our model in color-color space. Stars classified as BHBs, BSs, and outliers are shown in blue circles, purple squares, and red stars, respectively.  The predicted class is defined to be the one that shows the highest posterior-marginalized probability, and the color is used to represent the probability (one of $p_{\mathrm{BHB}}, p_{\mathrm{BS}}, p_{\mathrm{outlier}})$. The blue solid line and purple solid line show the ridgelines $\mu_{\mathrm{BHB}}, \mu_{\mathrm{BS}}$, calculated from the best-fit parameters in Table \ref{table:Mixture Model Prior}. The shaded regions with dashed lines on their boundaries along each ridgeline show the intrinsic scatters $\sigma_{\rm int}$ for each sequence.
Furthermore, the predicted ratios of BHB stars (blue), BS stars (purple), and outliers (red) as a function of magnitude are shown in the middle right of Figure \ref{fig:Mixture_Model}.

\begin{figure*}[!htb]
    \centering
    \includegraphics[width=0.98\textwidth]{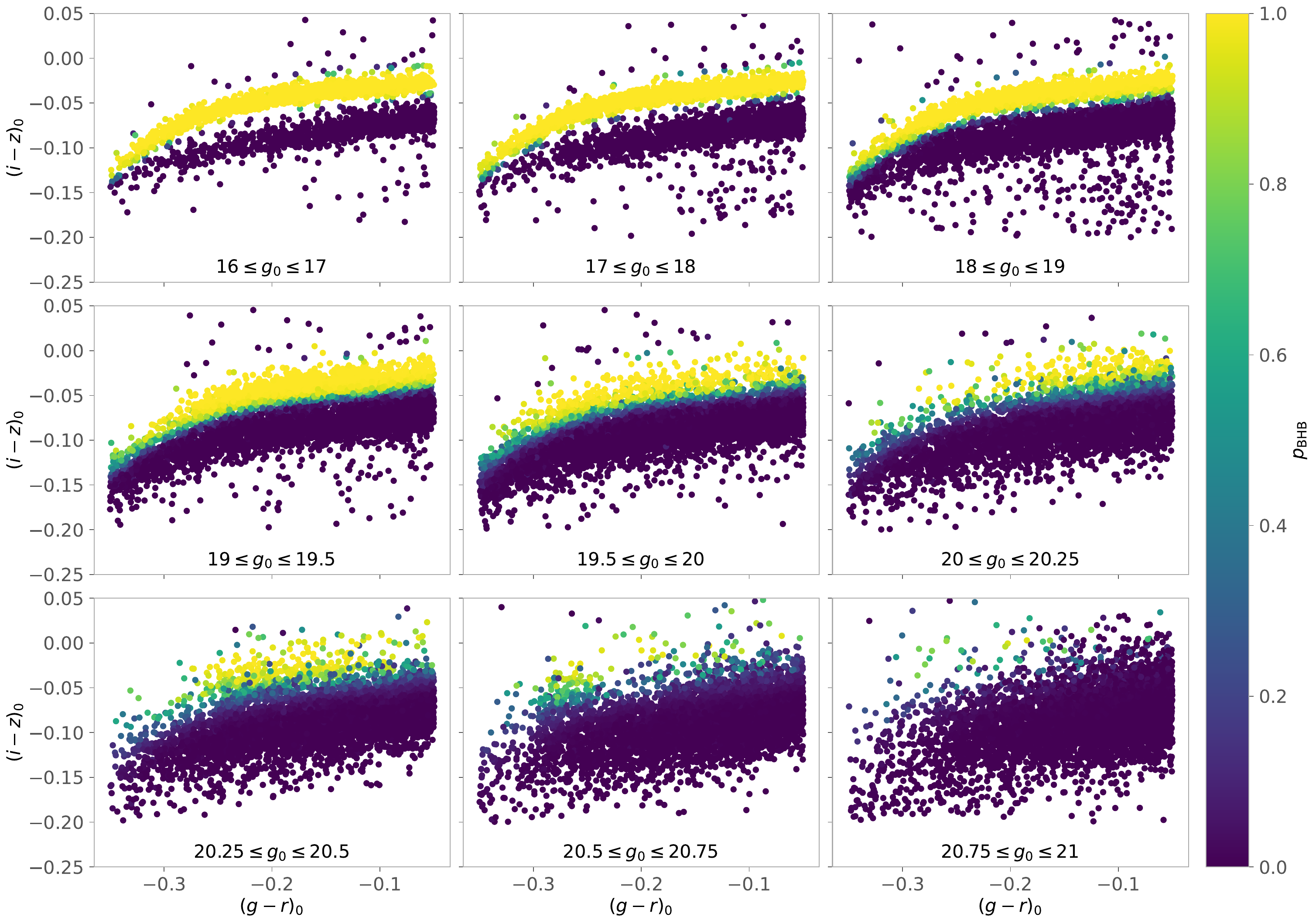}
    \caption{As Figure \ref{fig:BHB_magnitude}, but now highlighting the predicted BHB probability $p_\mathrm{BHB}$ for individual  stars as a function of $g_0$ magnitude. The model easily classifies BHBs at brighter magnitudes ($g_0 < 19$). The two sequences start to overlap at $19 < g_0 < 20$. At the faintest magnitudes, the changing relative fraction of BHB versus BS stars, in addition to larger photometric uncertainties, leads to a more ambiguous classification.
    }
    \label{fig:BHB_wrt_magnitude}
\end{figure*}

Using the mixture model, we predict the probabilities of being a BHB star from 16 mag to 21 mag. Figure \ref{fig:BHB_wrt_magnitude} shows stars from selected magnitude bins. The model distinguishes BHB versus BS easily at brighter magnitudes $(g_0 < 19)$ with a clear decision boundary. At fainter magnitudes, we see there are substantially fewer BHBs, because of the larger uncertainties and decreasing ratio of BHB versus BS at these fainter magnitudes.

\subsection{Validation from cross-matching with \SSSSS Data}
We apply our mixture model to \SSSSS x DES DR2 data to compute the BHB probability and use a threshold cutoff on the probability of $p_\mathrm{BHB} \geq 0.5$ to distinguish between BHB and BS stars. The resulting predictions are shown in the bottom right panel of Figure \ref{fig:photometry_comparison}. It matches the prediction from the spectroscopic parameters in \SSSSS well (see the bottom left panel), with recall 0.962, precision 0.953, and false positive rate 0.047, suggesting the robust performance of our model.

Using the predicted BHB probabilities, we select the BHB candidates from DES DR2 and use them for subsequent analysis to derive the density profile of the Milky Way.

\begin{figure*}[!htb]
    \centering
    \includegraphics[width=0.98\textwidth]{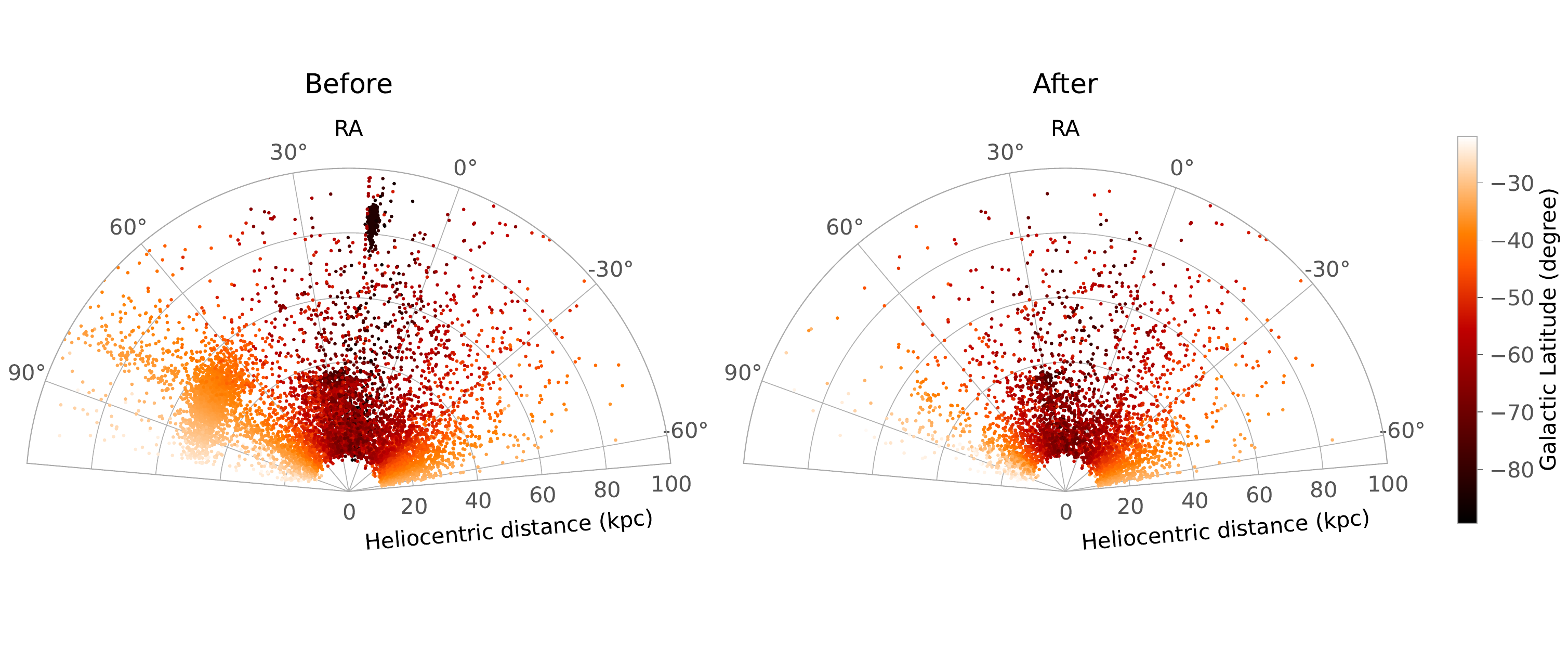}
    \caption{Radial distribution of BHB candidates (with $p_{\mathrm{BHB}} \geq 0.5$ computed by our model) selected from DES DR2. Left: all candidates. Right: after removing regions associated with known overdensities in Section \ref{sec:Density}. Color represents Galactic latitude b. In order of increasing heliocentric distance, our selected BHBs correctly show overdensities associated with known substructures: the Sgr stream, the LMC, and Sculptor. The removal is not exhaustive, and there are many other substructures left in the sample (see Figure \ref{fig: substructures}). 
    }
    \label{fig:Radial_density_map}
\end{figure*}

  \begin{figure*}[!htb]
    \centering    \includegraphics[width=0.98\textwidth]{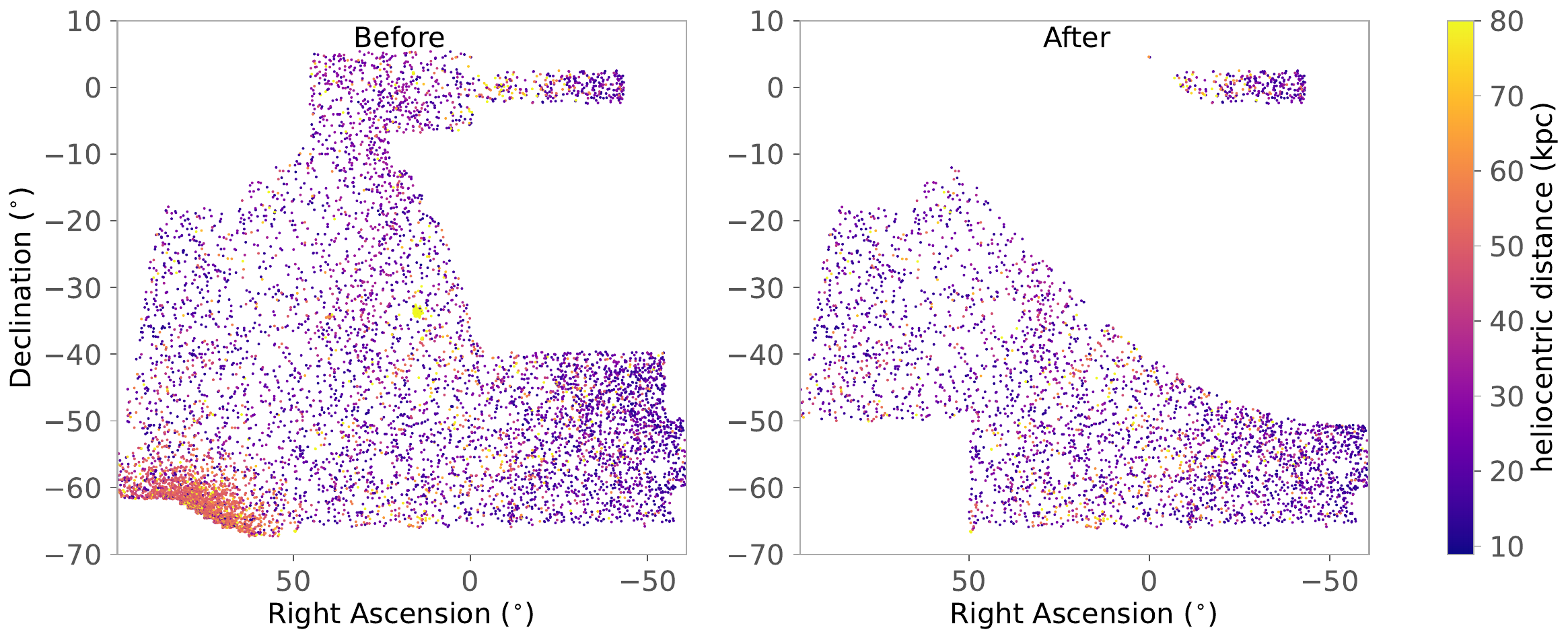}
    \caption{On-sky distribution of BHB candidates, for the same samples as in Figure \ref{fig:Radial_density_map}. Left: before removing substructures. Right: after removing substructures. Color represents heliocentric distances. The removal of regions associated with LMC, NGC300, Fornax, Sculptor, and Sgr are clearly shown.  
    }
    \label{fig:Density_Map}
\end{figure*}

\begin{figure*}[!hbt]
    \centering    \includegraphics[width=0.98\textwidth]{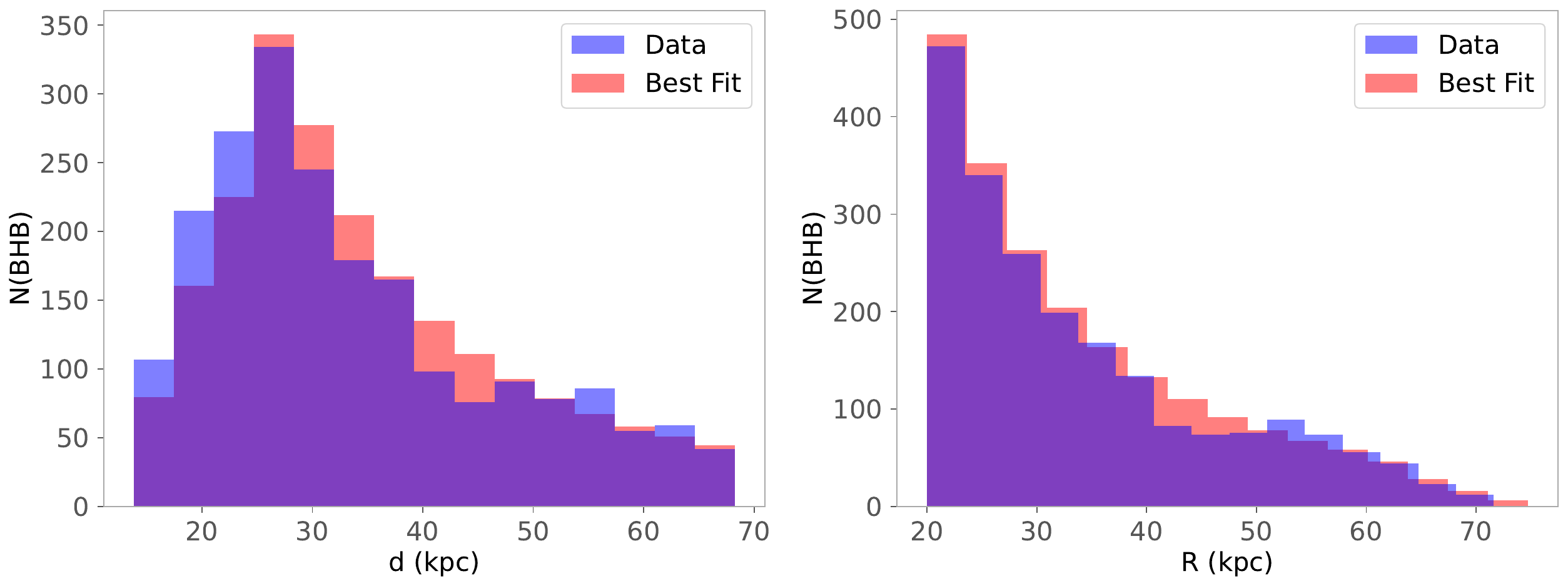}
    \caption{Histograms of final selected BHB candidates in DES DR2 versus simulated samples from the best-fit density profile. Left: with heliocentric distance $d$. Right: with Galactic radius $R$. We see that our best-fit density profile accurately captures the stellar density distribution, generating samples that match the data well from both heliocentric and Galactocentric points of view. 
    }
    \label{fig:BHB hist}
\end{figure*}

\section{Density Profile}\label{sec:Density}
\subsection{Calibration to Distance and Selection of BHBs}
We enforce a cut $p_{\mathrm{BHB}} \geq 0.5$ to select the BHB stars. The distances of BHBs can be calibrated in a straightforward manner. We employ the relations between absolute magnitude $M_g$ and $(g-r)_0$ color from \citet{2016MNRAS.456..602B},
\begin{multline}\label{eq:mg}
    M_g = 0.398 - 0.392(g-r)_0 + 2.729((g-r)_0)^2 \\ + 29.1128((g-r)_0)^3 + 113.569((g-r)_0)^4.
\end{multline}
\noindent See also a similar relation from  \citet{ 2011MNRAS.416.2903D}. We then use $g_0 - M_g$ to obtain the distance modulus, which is subsequently converted to heliocentric distance ($d$).
 
To characterize the underlying density profile of the halo itself, we need to remove stars associated with substructures that will likely result in overdensities. Known substructures in the Milky Way halo which are in the vicinity of our sample include Sculptor, Fornax, the Sgr stream, the Large Magellanic Cloud (LMC). Accordingly, stars in regions close to these substructures are excluded from our density profile study. In addition, we also exclude the region near an external galaxy NGC 300 as 
there is an excess of blue stars in the DES DR2 photometry in this region. 
For Fornax and Sculptor, we use a 3-deg radius cut centered at $(\mathrm{RA},\mathrm{Dec}) = (39.9583^\circ, -34.4997^\circ)$ and $(15.0183^\circ, -33.7186^\circ)$, respectively. For the Sgr stream, we first convert the Galactic coordinate system to the heliocentric spherical coordinate system defined by the orbit of the Sagittarius dwarf galaxy \citep{2003ApJ...599.1082M}, with Sgr longitude-like angle $\Lambda$ and Sgr latitude-like angle $B$. Stars within the region $| B | \leq 20^\circ$ are removed. For the LMC, we remove stars within the region $ \mathrm{Dec} \leq -50^\circ, 50^\circ \leq \mathrm{RA} \leq 150^\circ$. For NGC 300, we use a 0.5-deg radius cut centered at $(\mathrm{RA},\mathrm{Dec}) = (13.7229^\circ, -37.6844^\circ)$.

We show the radial distribution of the selected BHB stars in Figure \ref{fig:Radial_density_map}, both before and after the removal of these overdensities. In order of increasing heliocentric distance, overdensities associated with the Sgr stream, the LMC, and Sculptor are visible in the full sample. We also present a view of the on-sky location of our BHB candidates, before and after removal of overdensities, in Figure \ref{fig:Density_Map}. 
We note that the removal of overdensities is not exhaustive, and there are numerous stars from 
substructures which remain in the sample. We discuss these further in section \ref{sec:discuss}.

To establish an accurate density profile, it is important to account for potential partial selection effects for the BHB sample, especially at the faint and bright extremes. We assume that the BHB sample in DES is 100\% complete in the magnitude range of $16<g_0<20.5$. Since the absolute magnitude of BHBs $M_g$ is a function of $g-r$ color, we compute the maximum distance for a BHB at $g_{\mathrm{min}}=16$ in the color range of $-0.35<(g-r)_0<-0.05$ as the lower limit of our complete sample, $d_{\mathrm{min}}=13.05$ kpc. Similarly, We use the minimum distance for a BHB at $g_{\mathrm{max}}=20.5$ in the same color range as the upper limit of our complete sample, $d_{\mathrm{max}}=68.37$ kpc. We consider our sample to be complete between these lower and upper limits, and thus model the density profile only across this region.
We additionally exclude stars with Galactocentric distances $R <$ 20 kpc, as we notice those stars are likely highly incomplete, and do not model the density profile in this region.
We obtain a total of 2103 BHBs after these selections, and we show histograms (blue) with respect to heliocentric distance (d) and Galactocentric distances ($R$) in the left and right panels of Figure \ref{fig:BHB hist} respectively.

\subsection{Inhomogeneous Poisson Point Process Model}
As the selected BHB candidates are influenced by 1) the spatial coverage of the DES survey and 2) the limits on Galactocentric distances ($R$), they are not a true representation of the underlying stellar distributions. Hence when fitting the expected stellar distribution, we must consider these two selection effects. Below we present the details of modeling the stellar density profile based on the selection effects, and we model each star using an inhomogeneous Poisson point process (IPPP). This method has been successfully employed in previous studies conducted by \citet{bovy_spatial_2012}, \citet{xue_radial_2015}, and \citet{han_stellar_2022}, among others. 

We define the Poisson intensity function in the Galactic coordinate system $(l, b, d)$. The rate of finding a star can be written as 
\begin{multline}
    \lambda(l,b,d) = A \times (R+R_\mathrm{smooth})^{-\alpha} \times  \\
    |\mathbf{J}(l,b,d|R)| \times \mathbf{S_{\mathrm{distance}}}(d, R) \times \mathbf{S_{\mathrm{sky}}}(l, b) 
\end{multline}
where
\begin{itemize}
    \item $R$ is the Galactocentric radius, defined by $R = \sqrt{ R_0^2 +d^2 - 2R_0 d \cdot \cos(l) \cdot \cos(d)}$ and $R_0$ is the distance from the Sun to the Galactic Center, taken to be 8.3 kpc \citep{2017ApJ...837...30G}. 
    \item $R_{\mathrm{smooth}}$ is a smoothing radius to allow for integration down to 0, and it is set to 1 kpc.
    \item $\alpha$ is the slope for the power law. We do not look for a power law with a breaking radius near 20 kpc as this is close to our lower limit in distance \citep{medina_discovery_2018, watkins_substructure_2009}. 
    \item $\mathbf{S_{\mathrm{distance}}}(d, R)$ and $\mathbf{S_{\mathrm{sky}}}(l, b)$ are the distance selection function and spatial selection function based on the DES footprint, respectively.
    \item $|\mathbf{J}(l,b,d|R)|$ is the Jacobian term to account for coordinate transformation and
    $|\mathbf{J}(l,b,d|R)| = d^2 cos(b)$
    \item $A$ is a normalization coefficient. \\
\end{itemize}
We define our selection functions to be simple binary indicators as a function of $R$, $d$, $l$, and $b$, with
\begin{equation}
    \mathbf{S_{\mathrm{distance}}}(d, R) = 
    \begin{cases} 
     1 & \text{if 20 kpc } \leq R \text{ and } d_{\mathrm{min}} < d < d_{\mathrm{max}} \\
     0 & \text{else} 
    \end{cases}
\end{equation} 
and
\begin{equation}
    \mathbf{S_{\mathrm{sky}}}(l, b) = \mathbf{S}(\mathrm{RA},\mathrm{Dec}) =
    \begin{cases}
     1 & \text{if } -60^\circ \leq \mathrm{RA} \leq 100^\circ \\ \text{ and } & \text{   } -70^\circ \leq \mathrm{Dec} \leq 10^\circ\\
     0 & \text{else}
    \end{cases}
\end{equation}
which approximately constrains us to within the coverage of the DES footprint. Intuitively, $\lambda \cdot (dl \cdot db \cdot dd) $ can be interpreted as the infinitesimal probability of a star existing in an infinitesimal volume region $(dl, db, dd)$ located at position $(l, b, d)$.

With the rate parameter $\lambda$, the log-likelihood of an IPPP  can be written as
\begin{align}
    \log L &= \sum \log L(l,b,d | \boldsymbol{\phi}) \nonumber \\ &= -\Lambda + \sum \log\lambda(l,b,d | \boldsymbol{\phi})
\end{align}
where $\boldsymbol{\phi}$ represents the set of parameters $\alpha$ and $A$. $\Lambda$ is the integral of $\lambda$, and 
\begin{multline}
    \Lambda = A \, \int_{d_{\mathrm{min}}}^{d_{\mathrm{max}}} \int_{0}^{2 \pi} \int_{-\frac{\pi}{2}}^{\frac{\pi}{2}} d^2 \, \cos(b) \, (R+R_{\mathrm{smooth}})^{-\alpha} \times \\
    \mathbf{S_{\mathrm{distance}}}(d, R) \, \mathbf{S_{\mathrm{sky}}}(l, b) \, {\rm d}b \, {\rm d}l \, {\rm d}d
\end{multline}

\noindent As the integral $\Lambda$ is analytically intractable, we approximate it using the trapezoidal rule by integrating over grids of $l, b, d$. To speed up the computation, we specify a grid of $\alpha$ values ($\Lambda$ is a function of $\alpha$) and pre-computed the integrals at each $\alpha$. We then use interpolation from \texttt{scipy} \citep{2020SciPy-NMeth} on the $\alpha$ grid such that during the actual sampling process the integral can be easily obtained by inputting $\alpha$. We again assume uniform priors on the two parameters as listed in Table \ref{table:Density Profile Prior}. Similarly to Section \ref{sec:model_classification}, we sample from the posterior using the \texttt{dynesty} Nested Sampling package with the default settings.\\

\begin{table}
\begin{center}
\caption{
Uniform prior range, samples that has the
maximum likelihood (best-fit), and uncertainties of the 2 parameters in the density model.
}
\label{table:Density Profile Prior}
\begin{tabular}{c c c c c}
\hline
 & Prior & Best 
 & 68$\%$ Credible & 95$\%$ Credible\\
 Parameter  & Range & Fit &  Interval & Interval \\
\hline
\hline
        $\alpha$ & (2, 7) &  4.28 & ($-$0.07, 0.07) &  ($-$0.12,0.13) \\
        $\log A$ & (1,16) & 10.97 &  ($-$0.25, 0.24) &($-$0.44, 0.47) \\  
\hline
\end{tabular}
\end{center}
\end{table}

\begin{figure}[!htb]
    \centering    \includegraphics[width=0.48\textwidth]{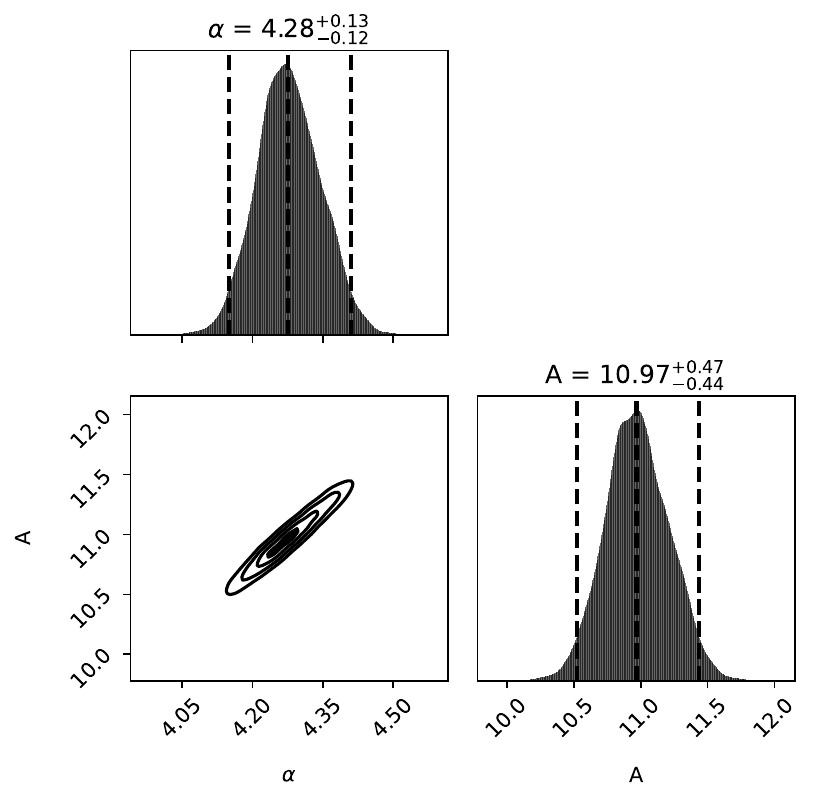}
    \caption{A corner plot showing the marginal posterior probability distributions of the two parameters of our density profile estimated with \texttt{dynesty}. We see that the power-law index $\alpha$ and normalization coefficient $A$ are well-constrained, and they are correctly positively correlated since the total stellar count is constant.}
    \label{fig: Corner plot}
\end{figure}

\begin{figure*}[!hbt]
    \centering    \includegraphics[width=1.0\textwidth]{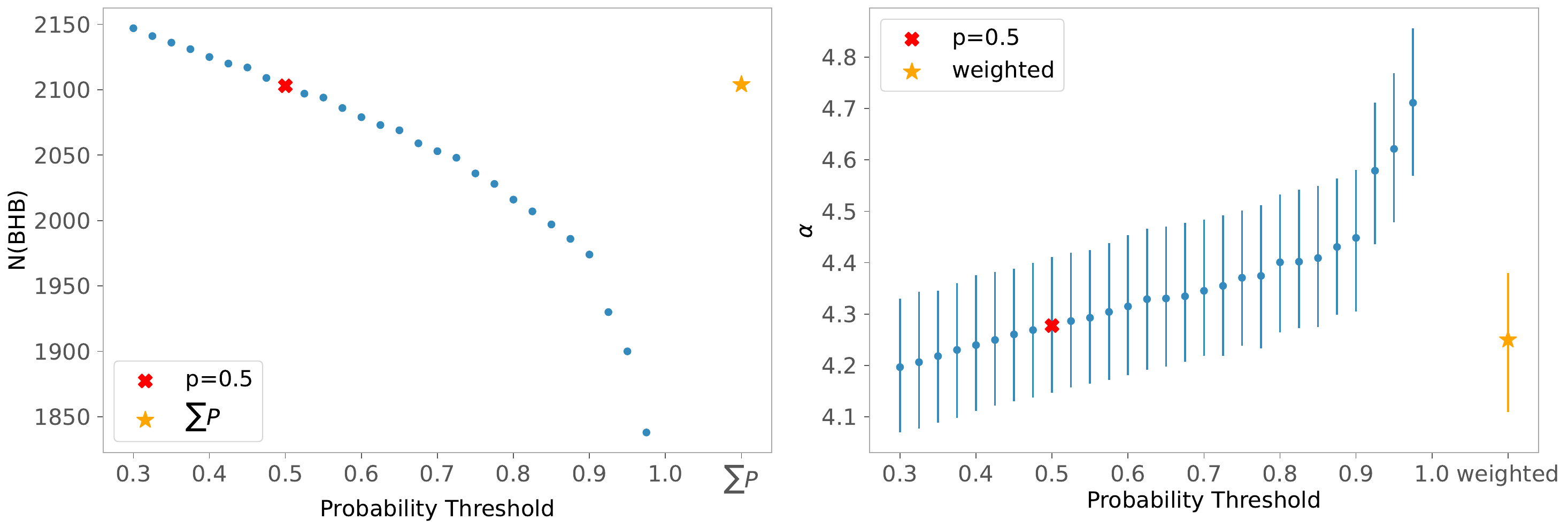}
    \caption {Left: BHB candidate count given different probability thresholds. Right: best-fit power-law index $\alpha$ for BHBs samples corresponding to the differing probability thresholds $p$. The error bar shows $95\%$ $ (2\sigma)$ credible interval.The normalization coefficient $A$ is completely degenerate with $\alpha$ and is thus omitted. The criterion we used for the previous section $p_\mathrm{BHB} \geq 0.5$ is highlighted. As the selection becomes stricter, BHB count decreases and we observe a steeper descent. We note that the trends in both panels are not smooth, with small bumps discernible, which is explained in Figure \ref{fig:average BHB probability}.
    The orange star shows the best-fit $\alpha$ where we weigh the log-likelihood of each star by its BHB probability, instead of using the unweighted log-likelihood of BHB candidates selected from hard-thresholding (see text for details). 
    }
    \label{fig:BHB threshold}
\end{figure*}

\begin{figure}[!hbt]
    \centering\includegraphics[width=0.48\textwidth]{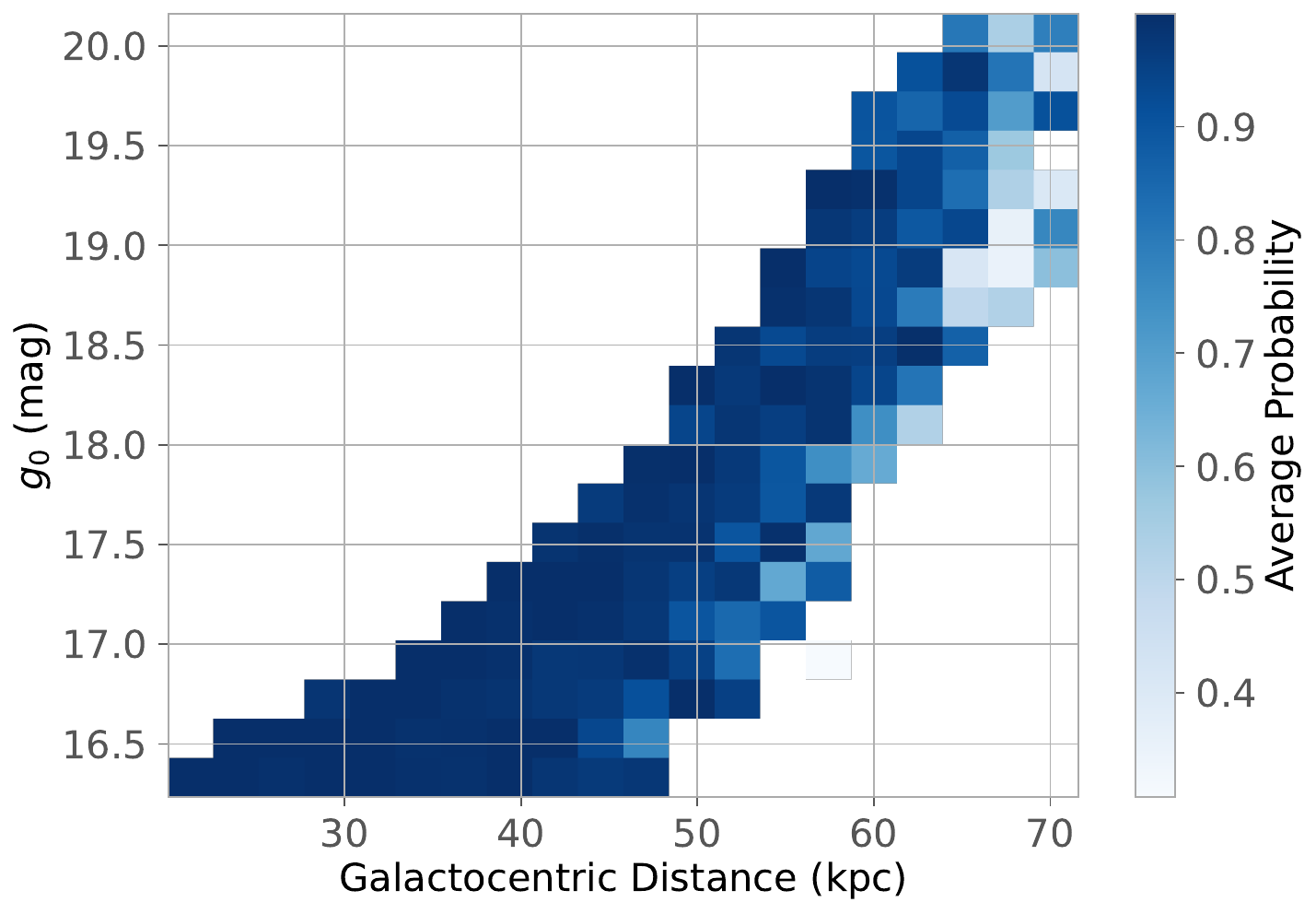}
    \caption{Averaged BHB probability with respect to $g_0$ band photometry and Galactocentric distance. Stars are selected with $p_\mathrm{BHB} \geq 0.3$ (same as the first threshold in Figure \ref{fig:BHB threshold}. Then the selected candidates are separated into different 2D bins based on $g_0$ and Galactocentric distance R. Within each bin, the average BHB probability is computed over the stars inside that bin, shown by color. We observe that BHB probability generally decreases as distance increases, or when the magnitude becomes larger. However, it is worth noting that the trend is not smooth with respect to magnitude and distance. Some bins appear to have higher BHB probabilities than their neighbours. A hard probability threshold cutoff may accidentally remove a group of stars in those bins, resulting in the bumps we see in Figure \ref{fig:BHB threshold}.
    }
    \label{fig:average BHB probability}
\end{figure}

\subsection{Results}
We find $\alpha$ that maximizes the likelihood and its $95\%$ $(2 \sigma)$ credible interval to be $4.28_{-0.12}^{+0.13}$ and normalization factor A to be $10.97_{-0.44}^{+0.47}$. In Figure \ref{fig: Corner plot}, 
the corner plot shows the 2D and 1D marginal posterior probability distributions of the two parameters of our density profile estimated with \texttt{dynesty}. We see that the power-law index $\alpha$ and normalization coefficient $A$ are well-constrained and positively correlated. This is expected given the constant stellar count $N (\rm{BHB})$.

\begin{figure*}[!hbt]
    \centering
    \includegraphics[width=0.98\textwidth]{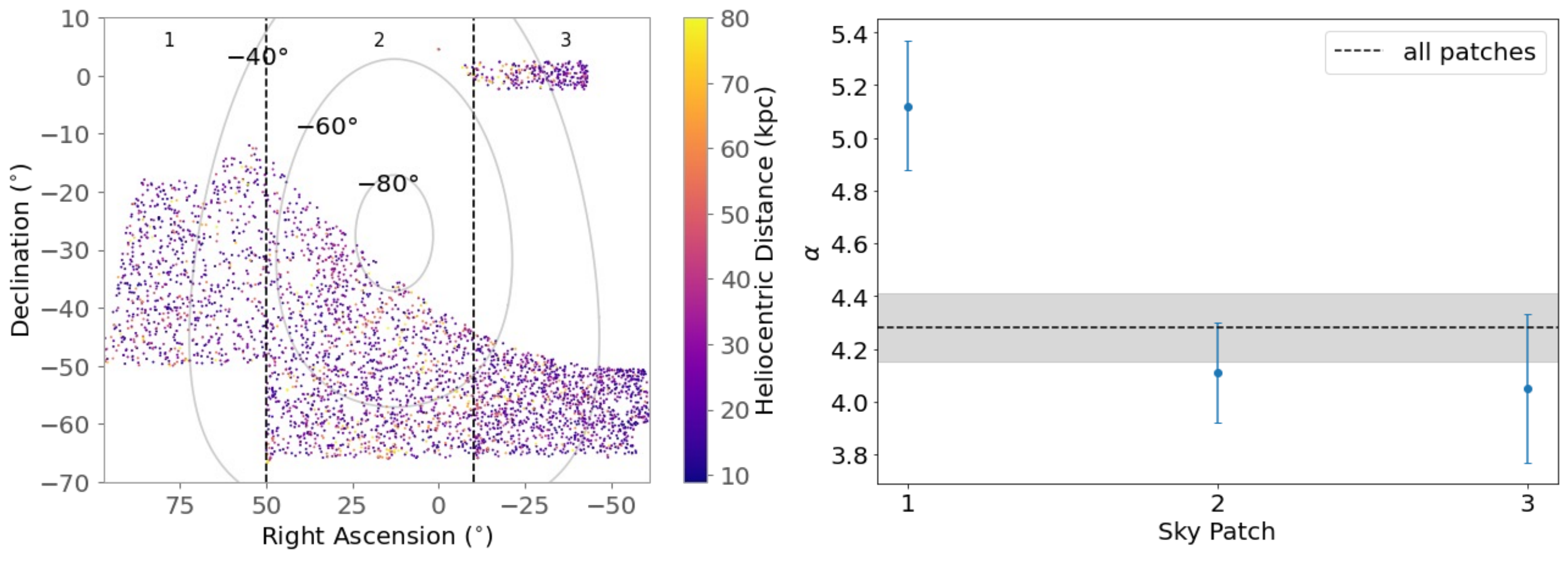}
    \caption{Left: We split DES coverage into 3 patches, shown by the dashed lines. Iso-latitude lines at $b$=$-$80$^\circ$, $-$60$^\circ$, $-$40$^\circ$ are shown in grey. Right: best-fit parameter $\alpha$ with $95\%$ (2$\sigma$) credible interval of the BHB density profile within each patch. The dashed line and shaded region show the best-fit parameter for the entire DES footprint. We see that $\alpha$ varies considerably among different sky patches, which reflects the spatial inhomogeneity of the Milky Way halo and illustrates the importance of considering the location and coverage of a survey when fitting a density profile.
    }
    \label{fig:sky patches}
\end{figure*}

\begin{figure*}[!hbt]
    \centering    \includegraphics[width=0.98\textwidth]{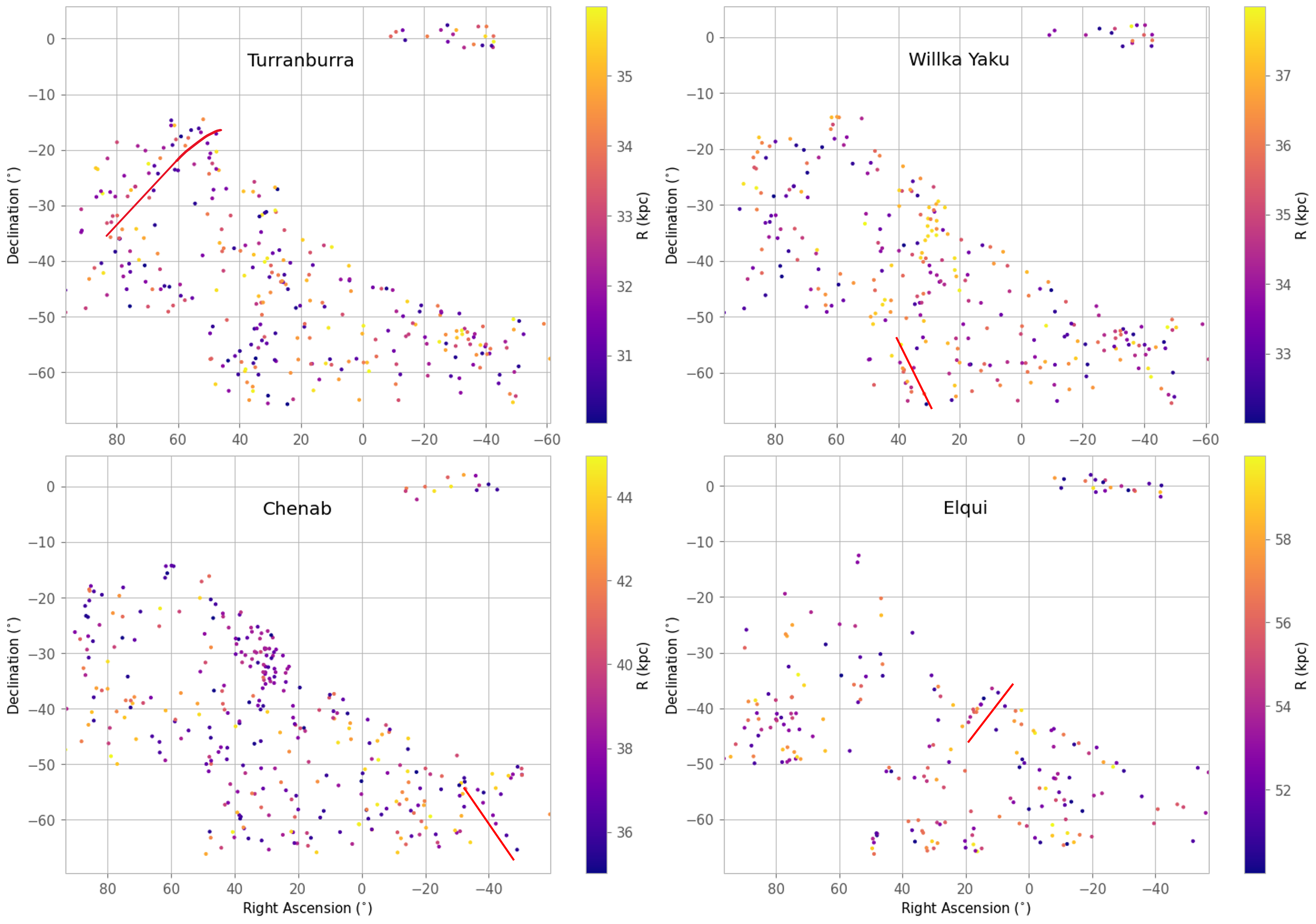}
    \caption{Similar to Figure \ref{fig:Density_Map}, but showing only BHB candidates within certain Galactocentric distances, as shown by the range of each colorbar. Moving from the top left to the bottom right, we present examples of four streams: Turranburra, Willka Yaku, Chenab, and Elqui, which are shown in red. Our sample displays overdensities that align closely with these previously identified substructures. This alignment demonstrates the consistency of our catalog with prior discoveries and underscores its potential utility in future searches for additional substructures. 
    }
    \label{fig: substructures}
\end{figure*}

\subsection{Mock Data}
As a verification, we generate mock data from our best-fit parameters (that maximize the likelihood) and compare the predicted stellar distribution with the actual distribution. To generate the predicted stellar distribution, we construct a grid of $l, b, d$ values. For each point $(l,b,d)$, we compute the density function $\lambda$ with the two selection functions $\mathbf{S}$, and record the distances $d, R$. Then, we use the trapezoidal rule over the grid to compute the integral of densities $\lambda$. We normalize this integral so it is equal to the total number of BHBs. After normalization, the previously computed density $\lambda$ at different distances $d, R$ will approximate the predicted stellar count. In Figure \ref{fig:BHB hist}, we compare the predicted stellar count (shown in red) to the observed distribution (shown in blue) for both the heliocentric distances $d$ (left panel) and Galactocentric distances $R$ (right panel). The distributions show good agreement, indicating that our model captures the stellar density distribution of the data well.

\section{Discussion}\label{sec:discuss}
\subsection{Systematic Error of Hard-Thresholding}
During the selection process of BHBs, we use a hard probability threshold cutoff to select sources with $p_\mathrm{BHB} \geq0.5$. Hard thresholding is likely to introduce artificial contamination to our BHB sample. Although a threshold of 0.5 appears to yield high accuracy for bright stars (shown by testing on S5 $\times$ DES DR2 in Figure \ref{fig:photometry_comparison}), it is likely to overestimate or underestimate the star count in the fainter range due to the increased uncertainty. Thus, it is essential to investigate the systematic error as a result of applying different threshold values.

To address this, we generate 28 evenly-spaced (spacing=0.25) probability threshold values ranging from $p_\mathrm{threshold}=0.3$ to $1.0$. Under each threshold, we select BHB candidates and conduct the same analysis in Section \ref{sec:Density} to estimate A and $\alpha$. The results are presented in Figure \ref{fig:BHB threshold}, with the BHB count and estimate of the power law index $\alpha$ displayed in the left and right panels, respectively. The normalization coefficient $A$ is completely degenerate with $\alpha$ (they collectively define the number of stars) and is thus omitted. As the threshold increases, the power-law index increases. This matches our expectation since a stricter threshold cutoff will result in fewer BHBs selected in the fainter magnitude, corresponding to a steeper decline in the density, and equivalently, a larger power law index.

\begin{figure*}[!hbt]
    \centering
    \includegraphics[width=1.0\textwidth]{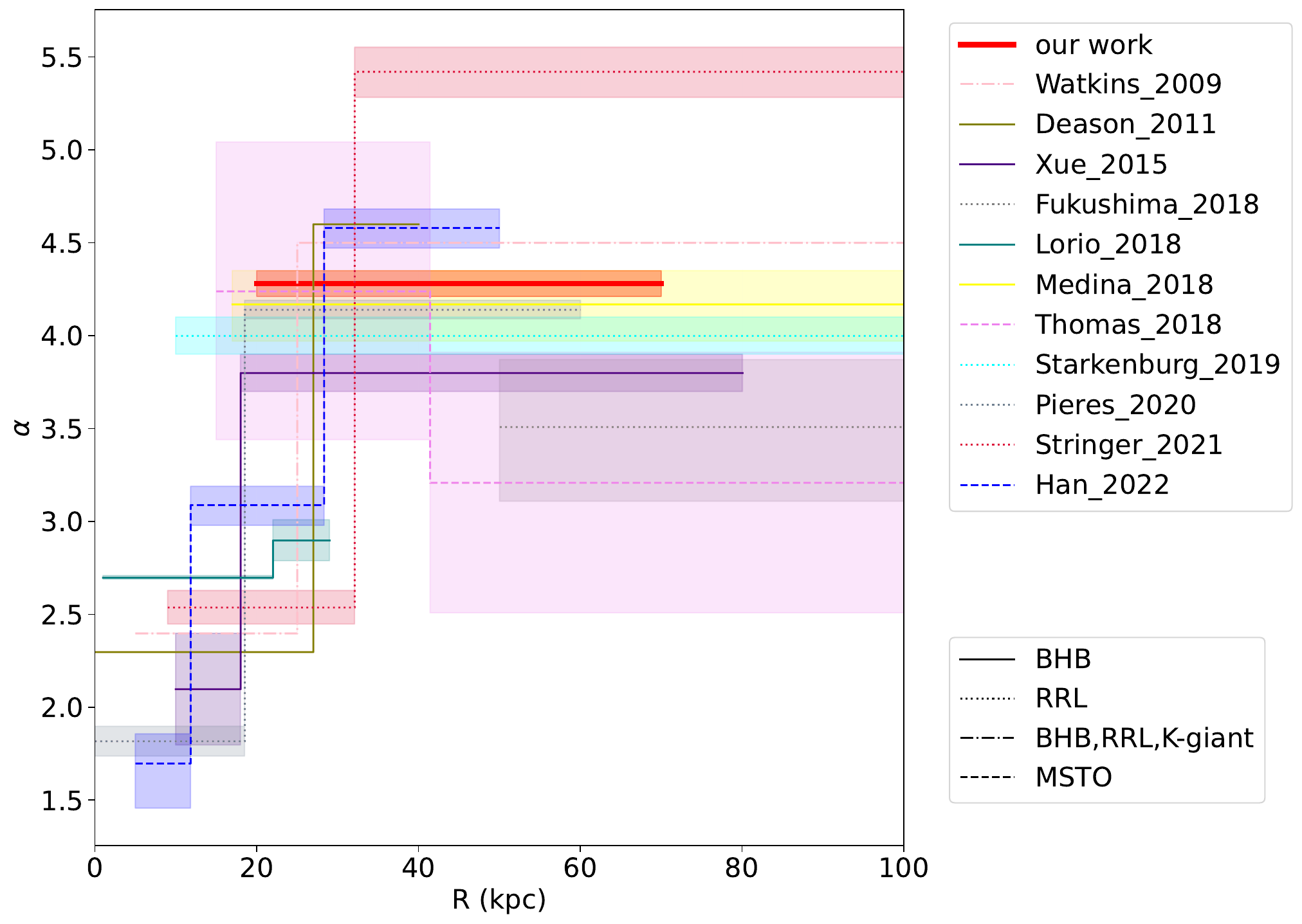}
    \caption{A comparison between our power-law index $\alpha$ (depicted in red) and values reported in existing literature \citep{ watkins_substructure_2009,deason_milky_2011,xue_radial_2015, fukushima_structure_2018,iorio_first_2018, medina_discovery_2018,thomas_-type_2018,   starkenburg_pristine_2019, Pieres_2020, Stringer_2021, han_stellar_2022}. The $\alpha$ values are plotted as a function of Galactocentric radius $R$, with shaded regions representing $68\%$ $(1 \sigma)$ credible interval, taken from corresponding literature when applicable. The lines are categorized based on the types of stars being analyzed. Within the distance range covered by our sample, our best-fit $\alpha$ closely matches the findings of previous studies.
    }
    \label{fig: literature comparison}
\end{figure*}

Notably, the plot reveals some small jumps in BHB count and power-law index as the threshold becomes stricter (at $p_\mathrm{threshold} \sim 0.9) $. We hypothesize this is likely due to the local overdensities in the halo, so that a specific threshold cutoff might accidentally remove a group of stars. To visualize this, we calculate the average BHB probability with respect to $g_0$ magnitude (as a proxy for heliocentric distance) and Galactocentric distance $R$. After selecting stars with $p_\mathrm{BHB} \geq 0.3$, which is the lowest threshold in Figure \ref{fig:BHB threshold}, we separate these candidates into different groups based on both $g_0$ band photometry and Galactocentric distance $R$. Within each group, the average BHB probability
is computed, shown as the value in each 2D bin in Figure $\ref{fig:average BHB probability}$. We see that the average BHB probability generally decreases as distance increases, or when the magnitude becomes larger. However, it is worth noting that the trend is not smooth with respect to magnitude and distance. There are some bins that have higher average BHB probabilities than their neighbours. A hard probability threshold cutoff may accidentally remove these small overdensities, resulting in the bumps seen in Figure \ref{fig:BHB threshold}.

The variations in the threshold value exert a non-negligible influence on the fitted parameters. This experiment assesses the systematic error and demonstrates that depending on the method, the true power-law index should fall between $4.2$ to $4.5$. 

Incorporating the probabilities as weights in the density model fitting process could potentially enhance the results. To this end, we explore another method where we drop the threshold cutoff completely and use the probabilities as weights in the log-likelihood for the sampling process instead. 

To compare with $N(\mathrm{BHB})$ from the thresholding approach, we sum up the BHB probabilities over all the sources (that possibly include BHBs, BSs, and outliers), shown as the orange star in the left panel of Figure \ref{fig:BHB threshold}. To compare with $\alpha$, we use these BHB probabilities to define a weighted log-likelihood function. The best-fit $\alpha$ that maximizes this likelihood is shown in orange in the right panel of Figure \ref{fig:BHB threshold}.  The weighted log-likelihood function multiplies the log-likelihood $\log \lambda$ of each source by its BHB probability. Using this approach, both the summed probability and power law index seem to align with N(BHB) and $\alpha$ from the 0.5 threshold cutoff. However, for the weighted log-likelihood approach, after we remove the log in the likelihood, multiplying by the weight effectively exponentiates the likelihood value of each star by its respective BHB probability, which lacks a clear physical meaning. Therefore, we include it for reference purposes but do not use it in our analysis. 

\subsection{Sky Variation} \label{sec:variation}
The stellar distribution can be influenced by the particular region of the sky under investigation. To understand the impact of different regions of the sky on the resulting density profile, we partition the DES footprint into three distinct patches, as illustrated in the left panel of Figure \ref{fig:sky patches}. We apply the same analysis in Section \ref{sec:Density} to each patch and find the power law indexes of those patches range from 4.05 to 5.12, presented in the right panel of Figure \ref{fig:sky patches}. The variation of power law indexes among different regions suggests that the Milky Way halo is spatially inhomogeneous, and that it is important to consider the location and coverage of a survey when fitting a density profile. In Section \ref{sec:limit}, we further discuss the implication of the sky variation on a flattened or triaxial halo density profile and suggestions for future investigations.

\subsection{Search for Substructures}\label{sec:substructures}
It is worth highlighting that the removal of substructures is not exhaustive during the data processing phase. As an example, when we select stars at specific distances of stellar streams discovered \citet{2018ApJ...862..114S}, we are able to recover the presence of several previously discovered thinner streams, including Turranburra,  Willka Yaku, Chenab, and Elqui, shown in Figure \ref{fig: substructures}. Our catalog thus assumes importance to complement the existing stellar stream catalog and proffer candidates for new streams.  

In addition, more diffused and massive streams like Palca/Cetus can be vaguely seen in Figure \ref{fig:Radial_density_map} and \ref{fig:Density_Map} around RA $\sim 30^\circ$. Inevitably, these substructures may also have some impact on the halo density measurement, which may also impact the sky variation discussed in Section \ref{sec:variation}. However, given the large number of streams in the survey area, it is not possible to remove {\it all} stream members.

\subsection{Comparison with Existing Literature}
We compare our estimated density profile power-law index $\alpha$ with existing literature in Figure $\ref{fig: literature comparison}$, including the measurements using photometric data from the DES (thus the same sky coverage), but with different tracers - RR Lyrae stars \citep{Stringer_2021} and main sequence turn-off (MSTO) stars \citep{Pieres_2020}. Our power law index is consistent with previous findings at the same Galactocentric distances $R$. 

Our investigations on different threshold cutoffs and different regions of the sky provide a good explanation for the variability in the existing literature. Models with different probability threshold cutoffs are analogous to the different methods used by other researchers, and we explain the variation for our model quantitatively by assessing the systematic error. The experiment on the different sky patches illustrates the dependence of the fitted density profile on the location and coverage of different surveys.

\subsection{Limitations and Future Studies}\label{sec:limit}
There are a number of ways we can expand upon the current study. Firstly, the analysis is conducted using stars within a distance range of 20 kpc to 70 kpc due to the conservative selection of high-probable stars. With future surveys with higher depth and better precision, we can map out more distant regions of the halo. 

Secondly, although our density model incorporates the fundamental characteristics of the stellar halo, it is limited by assuming a perfect spheroidal shape centered at the Galactic center, with an identical decline in all directions. 
However, the literature suggests that the true shape (and orientation) of the halo is more complex.
Several previous studies have claimed, for instance, that the halo is oblate \citep{Olling2000,Sesar2011,deason_milky_2011,Bowden2015,thomas_-type_2018}, prolate \citep{Helmi2004, Banerjee2011,Bowden2016,fukushima_structure_2018}, and triaxial \citep{Law2010,Deg2013, 2019MNRAS.482.3868I}, although spherical distributions \citep{Fellhauer2006,Smith2009,Das2023,Palau2023} and models where the halo flattening varies with Galactocentric radius or line of sight \citep[e.g.,][]{Vera-Ciro2013,hernitschek_profile_2018} have also been considered.
Additionally, it has been proposed that the halo likely exhibits a misalignment with respect to the galactic disk \citep[e.g.][]{han_stellar_2022}, as a result of its accretion history \citep[e.g.,][]{Prada2019,Dillamore2022}.
Furthermore, the halo density of the studied region may also be impacted by the wake induced by the infall of LMC \citep[][]{Belokurov2019, Conroy2021}. 
Indeed, the variation in $\alpha$ seen in Figure \ref{fig:sky patches} and Section \ref{sec:variation} may already imply that the density profile is more complicated than spherical, as significantly different declines in stellar densities are observed at similar Galactic latitudes. 
For future work, it would be advantageous to consider more flexible models that can accommodate such complexities. 

\subsubsection{Parametrization of the BHB/BS Mixture Model } 
In Section \ref{sec:model_classification} we parametrize our model that the ratio of BHB and BS is purely a function of $g_0$.
However, given that BHB and BS sequences represent two stellar populations with distinct evolutionary trajectories and distributions in the MW, this model might be too simplified. 

First, the ratio of the two populations is not expected to be constant as a function of $(g-r)_0$. For this, we experiment with a parametric form involving both $(g-r)_0$ and $g_0$ in the Appendix \ref{sec: complex parametrization}. As detailed in the Appendix, the resulting number of BHB stars as well as the power law index $\alpha$ with the new model shows little difference compared to the simpler model (i.e. without $(g-r)_0$ dependency).

Moreover, BHB stars are intrinsically brighter than BS stars, and thus locate much further than BS stars at the same apparent magnitude. Hence we also expect to see a wide range of observed BHB to BS ratios across different lines of sight, in particular, as a function of Galactic latitude where BS is likely dominated by the disk stars. However, since our model considers the sky as a whole, averaging out these variations, the differences in ratios along various lines of sight become less significant. For future research, it would be beneficial to model them as separate populations and parameterize densities of BHB and BS as functions of both distance and location on the sky (in Galactic latitude and longitude).

\section{Conclusions}\label{sec:conclusion}
In this study, we have developed a mixture model that predicts the probability of a star being a BHB based on its $g_0$ band magnitude, $(g-r)_0$ and $(i-z)_0$ photometry using DES DR2. Our study demonstrates that, even in the absence of the $u-$band, we can distinguish between BHB and BS sequences through precise $griz$ photometry. We identify $\sim2100$ highly probable BHB candidates in the Southern Hemisphere, and investigate the stellar halo within a distance range of 20 to 70 kpc. 
After excluding stars in the area associated with major known substructures, we observe a smooth decline in the stellar density, with a power law index $\alpha=4.28_{-0.12}^{+0.13}$, consistent with existing literature values. By drawing connections to our assessment of systematic error in threshold cutoffs and sky locations, we argue that the variations in current power law indexes in the literature can be largely associated with (a) different methodologies used to derive the density profiles, and (b) the inherent spatial inhomogeneity of the halo. We provide the entire catalog (which contains 46031 sources) with computed $p_\mathrm{BHB}$ using our model in Appendix \ref{sec: catalog}. We hope this catalog will be useful for future research on the Galactic halo.

To gain a more comprehensive understanding of the stellar halo, future studies should focus on using more flexible models and probing more distant stars in the halo. Such a BHB sample would not only help us study the stellar density profile of the Milky Way halo, but would also help identify old substructures such as stellar streams. The latter will contribute to our understanding of the Milky Way's accretion history. 

With forthcoming photometric surveys, like the Rubin Observatory's Legacy Survey of Space and Time (LSST), we anticipate not only extending our reach to greater distances but also expanding our observational scope across a broader expanse of the sky. This will enable us to investigate the density profile of the halo out to the virial radius of the MW and explore potential spatial inhomogeneity more comprehensively.
Overall, this study provides insights into the properties of the stellar halo and sets the stage for future investigations that aim to unravel the complex formation and evolution processes of our Galaxy.

\acknowledgments
The authors would like to thank the Summer Undergraduate Data Science (SUDS) Opportunities Program at the Data Science Institute at the University of Toronto for providing the funding and opportunity to enable this project.
T.S.L. acknowledges financial support from Natural Sciences and Engineering Research Council of Canada (NSERC) through grant RGPIN-2022-04794.
S.K. acknowledges support from Science \& Technology
Facilities Council (STFC) (grant ST/Y001001/1).
G.L. acknowledges FAPESP (proc. 2021/10429-0).
G.M.E. acknowledges financial support from NSERC through grant RGPIN-2020-04554.

The authors would like to thank Alex Drlica-Wagner, Andrew Pace, Adriano Pieres for their helpful comments.
This paper includes data obtained with the Anglo-Australian Telescope in Australia. We acknowledge the traditional owners of the land on which the AAT stands, the Gamilaraay people, and pay our respects to elders past and present.

For the purpose of open access, the author has applied a Creative
Commons Attribution (CC BY) licence to any Author Accepted Manuscript version
arising from this submission.

This research has made use of the SIMBAD database, operated at CDS, Strasbourg, France \citep{Simbad}.
This research has made use of NASA’s Astrophysics Data System Bibliographic Services.

This paper made use of the Whole Sky Database (wsdb) created by Sergey Koposov and maintained at the Institute of Astronomy, Cambridge by Sergey Koposov, Vasily Belokurov and Wyn Evans with financial support from the Science \& Technology Facilities Council (STFC) and the European Research Council (ERC).

This project used public archival data from the Dark Energy Survey
(DES). Funding for the DES Projects has been provided by the
U.S. Department of Energy, the U.S. National Science Foundation, the
Ministry of Science and Education of Spain, the Science and Technology
Facilities Council of the United Kingdom, the Higher Education Funding
Council for England, the National Center for Supercomputing
Applications at the University of Illinois at Urbana-Champaign, the
Kavli Institute of Cosmological Physics at the University of Chicago,
the Center for Cosmology and Astro-Particle Physics at the Ohio State
University, the Mitchell Institute for Fundamental Physics and
Astronomy at Texas A\&M University, Financiadora de Estudos e
Projetos, Funda{\c c}{\~a}o Carlos Chagas Filho de Amparo {\`a}
Pesquisa do Estado do Rio de Janeiro, Conselho Nacional de
Desenvolvimento Cient{\'i}fico e Tecnol{\'o}gico and the
Minist{\'e}rio da Ci{\^e}ncia, Tecnologia e Inova{\c c}{\~a}o, the
Deutsche Forschungsgemeinschaft, and the Collaborating Institutions in
the Dark Energy Survey.  The Collaborating Institutions are Argonne
National Laboratory, the University of California at Santa Cruz, the
University of Cambridge, Centro de Investigaciones Energ{\'e}ticas,
Medioambientales y Tecnol{\'o}gicas-Madrid, the University of Chicago,
University College London, the DES-Brazil Consortium, the University
of Edinburgh, the Eidgen{\"o}ssische Technische Hochschule (ETH)
Z{\"u}rich, Fermi National Accelerator Laboratory, the University of
Illinois at Urbana-Champaign, the Institut de Ci{\`e}ncies de l'Espai
(IEEC/CSIC), the Institut de F{\'i}sica d'Altes Energies, Lawrence
Berkeley National Laboratory, the Ludwig-Maximilians Universit{\"a}t
M{\"u}nchen and the associated Excellence Cluster Universe, the
University of Michigan, the National Optical Astronomy Observatory,
the University of Nottingham, The Ohio State University, the OzDES
Membership Consortium, the University of Pennsylvania, the University
of Portsmouth, SLAC National Accelerator Laboratory, Stanford
University, the University of Sussex, and Texas A\&M University.
Based in part on observations at Cerro Tololo Inter-American
Observatory, National Optical Astronomy Observatory, which is operated
by the Association of Universities for Research in Astronomy (AURA)
under a cooperative agreement with the National Science Foundation.

The Legacy Surveys consist of three individual and complementary
projects: the Dark Energy Camera Legacy Survey (DECaLS; NOAO Proposal
ID \# 2014B-0404; PIs: David Schlegel and Arjun Dey), the
Beijing-Arizona Sky Survey (BASS; NOAO Proposal ID \# 2015A-0801; PIs:
Zhou Xu and Xiaohui Fan), and the Mayall z-band Legacy Survey (MzLS;
NOAO Proposal ID \# 2016A-0453; PI: Arjun Dey). DECaLS, BASS and MzLS
together include data obtained, respectively, at the Blanco telescope,
Cerro Tololo Inter-American Observatory, National Optical Astronomy
Observatory (NOAO); the Bok telescope, Steward Observatory, University
of Arizona; and the Mayall telescope, Kitt Peak National Observatory,
NOAO. The Legacy Surveys project is honored to be permitted to conduct
astronomical research on Iolkam Du'ag (Kitt Peak), a mountain with
particular significance to the Tohono O'odham Nation.

NOAO is operated by the Association of Universities for Research in
Astronomy (AURA) under a cooperative agreement with the National
Science Foundation.

The Legacy Survey team makes use of data products from the Near-Earth
Object Wide-field Infrared Survey Explorer (NEOWISE), which is a
project of the Jet Propulsion Laboratory/California Institute of
Technology. NEOWISE is funded by the National Aeronautics and Space
Administration.

The Legacy Surveys imaging of the DESI footprint is supported by the
Director, Office of Science, Office of High Energy Physics of the
U.S. Department of Energy under Contract No. DE-AC02-05CH1123, by the
National Energy Research Scientific Computing Center, a DOE Office of
Science User Facility under the same contract; and by the
U.S. National Science Foundation, Division of Astronomical Sciences
under Contract No. AST-0950945 to NOAO.


{\it Software:} 
{\code{numpy} \citep{numpy}, 
\code{scipy} \citep{2020SciPy-NMeth},
\code{matplotlib} \citep{matplotlib}, 
\code{astropy} \citep{astropy,astropy:2018, astropy:2022},
\code{dynesty} \citep{2020MNRAS.493.3132S,koposov_joshspeagledynesty_2023},
}

\appendix
\begin{figure*}[!htb]
    \centering
    \includegraphics[width=0.98\textwidth]{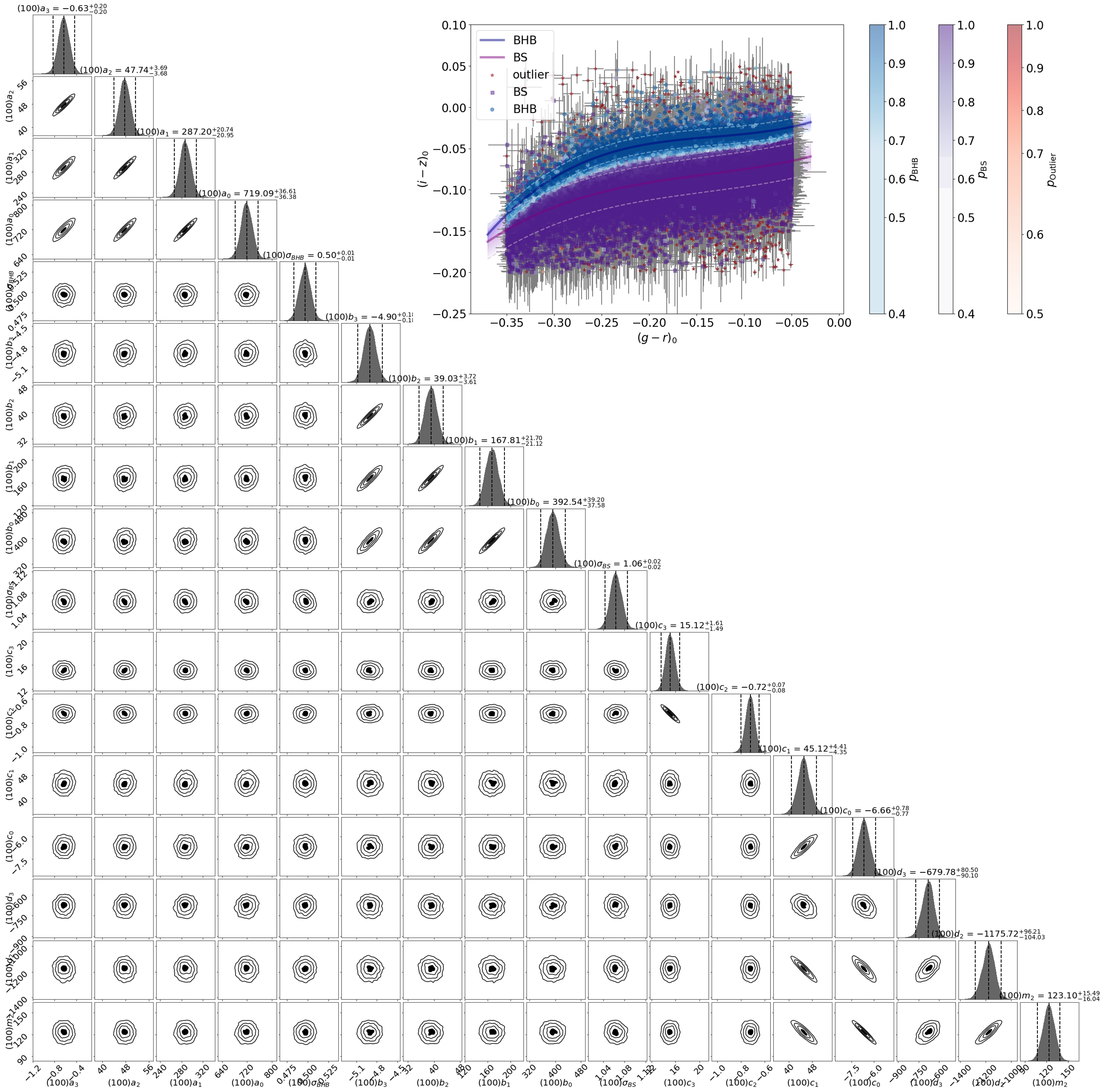}
    \caption{Same as Figure \ref{fig:Mixture_Model}, but with a more complex parametric form of the BHB ratio. Bottom left: A corner plot depicting the 2D and 1-D marginal posterior probability distributions of the 17 parameters of our photometric mixture model estimated with \texttt{dynesty}. \textit{All values are multiplied by a factor of 100 for visualization purposes.} The parameters $d_i$s are defined according to section \ref{sec: complex parametrization}. Top right: The predicted class probabilities for the stars used to derive the model in color-color space. BHB (blue) and BS (purple) sequences are shown as a solid line with the best-fit scatter as a light-shaded region with dashed lines defining its boundary. Stars classified as BHBs, BSs, and outliers based on their posterior-marginalized class probabilities are shown in blue circles, purple squares, and red stars respectively. The parameters are well-constrained, accurately trace the photometric distribution of stars, and successfully identify photometric outliers. }
    \label{fig:Complex_Mixture_Model}
\end{figure*}

\begin{figure*}[!htb]
    \centering
    \includegraphics[width=0.98\textwidth]{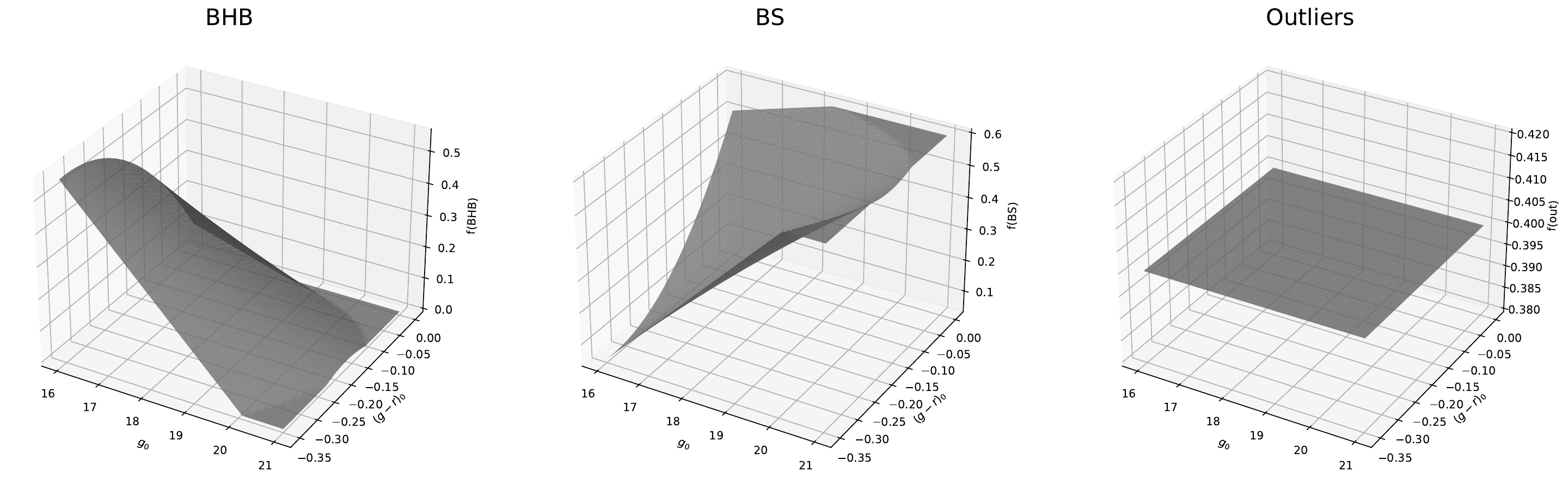}
    \caption{Same as bottom right of Figure \ref{fig:Mixture_Model}, but with a more complex parametric form of the BHB ratio. The predicted probability of a source in the BHB group, the BS group, and the outlier group, parametrized by equations defined in section \ref{sec: complex parametrization}. Our model generates physically-sensible magnitude and color dependencies across each subgroup. }
    \label{fig:Complex_model_ratio}
\end{figure*}

\begin{figure*}[!hbt]
    \centering    \includegraphics[width=0.98\textwidth]{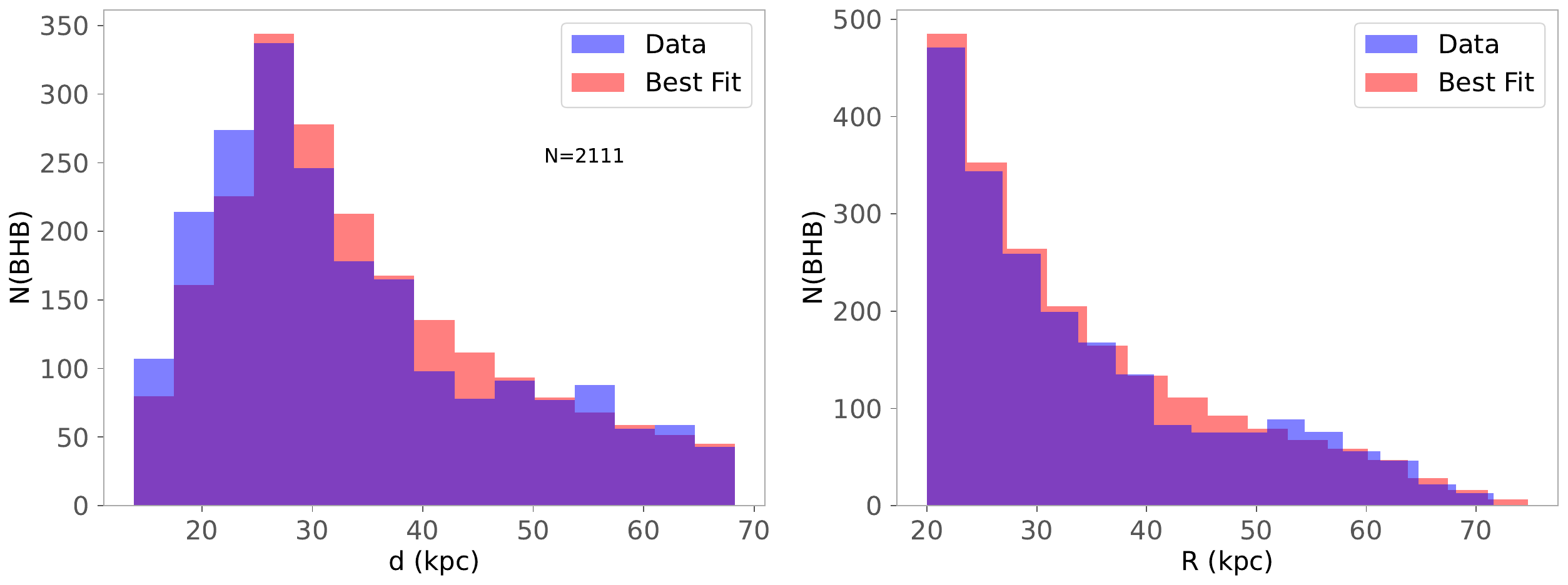}
    \caption{Same as Figure \ref{fig:BHB hist}, but with a more complex parametric form of the BHB ratio. Histograms of final selected BHB candidates in DES DR2 versus simulated samples from the best-fit density profile. Left: with heliocentric distance $d$. Right: with Galactic radius $R$. The number of BHBs is 2111, which is very close to 2103 in the previous fit.
    }
    \label{fig:complex BHB hist}
\end{figure*}

\begin{figure}[!htb]
    \centering    \includegraphics[width=0.48\textwidth]{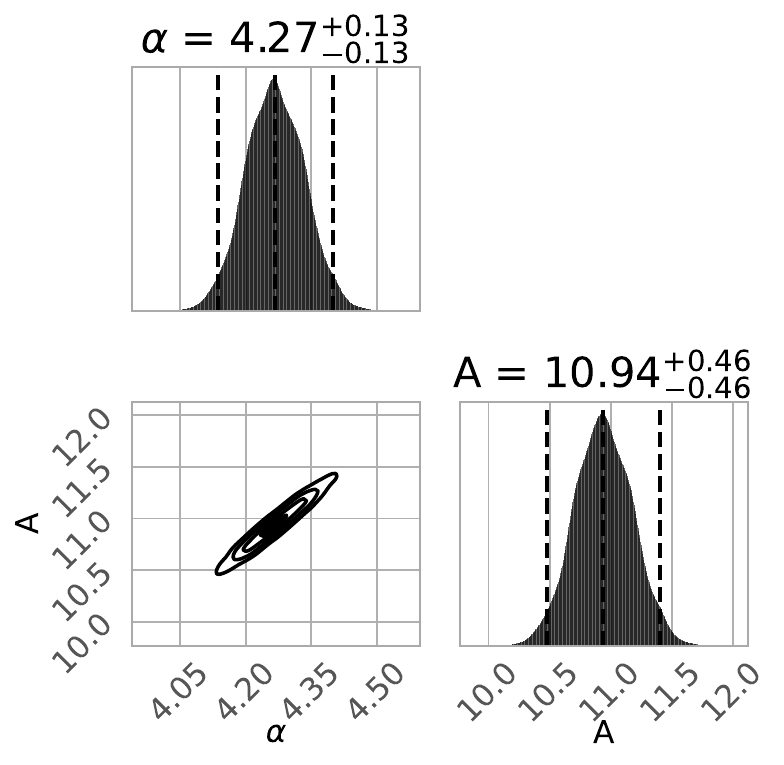}
    \caption{Same as Figure \ref{fig: Corner plot}, but with a more complex parametric form of the BHB ratio. A corner plot showing the marginal posterior probability distributions of the two parameters of our density profile estimated with \texttt{dynesty}. We see that the power-law index $\alpha = 4.27 ^{+0.13} _{-0.13}$, very close to the previous fit $\alpha = 4.28 ^{+0.13} _{-0.12}$.
    }
    \label{fig: complex corner plot}
\end{figure}

\section{BHB ratio as a function of both color and magnitude} \label{sec: complex parametrization}

We explore a parametric form of BHB ratio defined with $(g-r)_0$ and $g_0$. The statistical model for this parametric form is defined similarly as the previous (see Section \ref{subsec:like}), except that we describe the probability of having a source belong to the BHB group using a polynomial function of $g_0$ magnitude and $(g-r)_0$ color (vs equation \ref{eq: BHB ratio}, which is defined solely on $g_0$). We keep the probability of having a source belong to the outlier group as a linear function of $g_0$. This gives us five parameters:
\begin{multline}
    f_{\rm BHB}(g_0) = (1 - f_{\rm out}(g_0)) \times [d_0 + d_1 \times g_0 \\
    + d_2 \times (g-r)_0 +
    d_3 \times g_0 \times (g-r)_0 + d_4 \times (g-r)_0^2]
\end{multline}
with the probability of being in the BS class determined via:
\begin{equation}
    f_{\rm BS}(g_0) = 1 - f_{\rm BHB}(g_0) - f_{\rm out}(g_0)
\end{equation}
Hence we have a total of 17 unique parameters in our model to parametrize the distribution of BHB, BS, and outlier sources.

We conduct the same sampling process to find the parameters that maximize the likelihood, shown in Figure \ref{fig:Complex_Mixture_Model}. We also plot the probability of a source being each of the group, shown in Figure \ref{fig:Complex_model_ratio}. We then find the number of BHBs following the same processing discussed in Section \ref{sec:Density}. The number of BHB is found to be 2111, and the power law index $\alpha$ is $4.27 ^{+0.13} _{-0.13}$, shown in Figure \ref{fig:complex BHB hist} and Figure \ref{fig: complex corner plot}, respectively. Both of them show little difference from the previous fit.

\section{BHB catalog} \label{sec: catalog}

\begin{table*}
\centering
\caption{
A catalog with BHB probability $p_\mathrm{BHB}$ computed from the mixture model described in Section \ref{sec:model_classification} using $griz$ photometry from DES DR2. Distance is the heliocentric distance, computed assuming the sources are BHBs. A total of 46031 sources are included. Only the first ten lines are shown. The complete table is available online in a machine readable format.
}
\label{table:catalog}
\begin{tabular}{rrrcccccc}
\hline
Coadd Object ID & RA & Dec & $g_0$ & $r_0$ & $i_0$ & $z_0$ & distance & $p_\mathrm{BHB}$ \\
\hline
\hline
994953737 & 345.969096 & -42.549333 & 20.89 & 21.08 & 21.26 & 21.35 & 118.9 & 0.00 \\
999753919 & 346.598197 & -42.483844 & 19.32 & 19.54 & 19.73 & 19.78 & 56.18 & 0.95 \\
999777071 & 346.818066 & -42.739140 & 19.98 & 20.14 & 20.3 & 20.33 & 79.45 & 0.97 \\
995168807 & 348.283024 & -43.375280 & 17.19 & 17.50 & 17.74 & 17.84 & 17.53 & 0.98 \\
998800330 & 347.896323 & -43.028549 & 17.66 & 17.89 & 18.09 & 18.19 & 25.66 & 0.00 \\
1001308502 & 349.326354 & -43.156899 & 17.41 & 17.73 & 17.93 & 18.07 & 18.91 & 0.00 \\
998795299 & 348.005177 & -42.971495 & 20.24 & 20.36 & 20.50 & 20.60 & 90.63 & 0.00 \\
998786086 & 347.917693 & -42.875529 & 18.55 & 18.69 & 18.85 & 18.93 & 41.12 & 0.00 \\
998773023 & 347.836941 & -42.732092 & 17.52 & 17.66 & 17.82 & 17.89 & 25.72 & 0.00 \\
1001283704 & 348.961490 & -42.916494 & 18.60 & 18.65 & 18.75 & 18.80 & 43.20 & 0.00 \\
\hline
\end{tabular}
\end{table*}
Table \ref{table:catalog} shows a compiled catalog from Section \ref{sec:dataprep}, which contains 46031 sources, with $p_\mathrm{BHB}$ computed by our mixture model described in Section \ref{sec:model_classification} and heliocentric distance $d$ computed assuming the sources are BHBs using Equation \ref{eq:mg}. We also provide Coadd Object ID, RA, Dec, and $griz$ photometry, obtained from DES DR2 for each source. 

\bibliography{main}{}

\begin{thebibliography}{}
\expandafter\ifx\csname natexlab\endcsname\relax\def\natexlab#1{#1}\fi
\providecommand{\url}[1]{\href{#1}{#1}}
\providecommand{\dodoi}[1]{doi:~\href{http://doi.org/#1}{\nolinkurl{#1}}}
\providecommand{\doeprint}[1]{\href{http://ascl.net/#1}{\nolinkurl{http://ascl.net/#1}}}
\providecommand{\doarXiv}[1]{\href{https://arxiv.org/abs/#1}{\nolinkurl{https://arxiv.org/abs/#1}}}

\bibitem[{{Abbott} {et~al.}(2018){Abbott}, {Abdalla}, {Allam}, {Amara},
  {Annis}, {Asorey}, {Avila}, {Ballester}, {Banerji}, {Barkhouse}, {Baruah},
  {Baumer}, {Bechtol}, {Becker}, {Benoit-L{\'e}vy}, {Bernstein}, {Bertin},
  {Blazek}, {Bocquet}, {Brooks}, {Brout}, {Buckley-Geer}, {Burke}, {Busti},
  {Campisano}, {Cardiel-Sas}, {Carnero Rosell}, {Carrasco Kind}, {Carretero},
  {Castander}, {Cawthon}, {Chang}, {Chen}, {Conselice}, {Costa}, {Crocce},
  {Cunha}, {D'Andrea}, {da Costa}, {Das}, {Daues}, {Davis}, {Davis}, {De
  Vicente}, {DePoy}, {DeRose}, {Desai}, {Diehl}, {Dietrich}, {Dodelson},
  {Doel}, {Drlica-Wagner}, {Eifler}, {Elliott}, {Evrard}, {Farahi}, {Fausti
  Neto}, {Fernandez}, {Finley}, {Flaugher}, {Foley}, {Fosalba}, {Friedel},
  {Frieman}, {Garc{\'\i}a-Bellido}, {Gaztanaga}, {Gerdes}, {Giannantonio},
  {Gill}, {Glazebrook}, {Goldstein}, {Gower}, {Gruen}, {Gruendl}, {Gschwend},
  {Gupta}, {Gutierrez}, {Hamilton}, {Hartley}, {Hinton}, {Hislop}, {Hollowood},
  {Honscheid}, {Hoyle}, {Huterer}, {Jain}, {James}, {Jeltema}, {Johnson},
  {Johnson}, {Kacprzak}, {Kent}, {Khullar}, {Klein}, {Kovacs}, {Koziol},
  {Krause}, {Kremin}, {Kron}, {Kuehn}, {Kuhlmann}, {Kuropatkin}, {Lahav},
  {Lasker}, {Li}, {Li}, {Liddle}, {Lima}, {Lin}, {L{\'o}pez-Reyes}, {MacCrann},
  {Maia}, {Maloney}, {Manera}, {March}, {Marriner}, {Marshall}, {Martini},
  {McClintock}, {McKay}, {McMahon}, {Melchior}, {Menanteau}, {Miller},
  {Miquel}, {Mohr}, {Morganson}, {Mould}, {Neilsen}, {Nichol}, {Nogueira},
  {Nord}, {Nugent}, {Nunes}, {Ogando}, {Old}, {Pace}, {Palmese},
  {Paz-Chinch{\'o}n}, {Peiris}, {Percival}, {Petravick}, {Plazas}, {Poh},
  {Pond}, {Porredon}, {Pujol}, {Refregier}, {Reil}, {Ricker}, {Rollins},
  {Romer}, {Roodman}, {Rooney}, {Ross}, {Rykoff}, {Sako}, {Sanchez}, {Sanchez},
  {Santiago}, {Saro}, {Scarpine}, {Scolnic}, {Serrano}, {Sevilla-Noarbe},
  {Sheldon}, {Shipp}, {Silveira}, {Smith}, {Smith}, {Smith}, {Soares-Santos},
  {Sobreira}, {Song}, {Stebbins}, {Suchyta}, {Sullivan}, {Swanson}, {Tarle},
  {Thaler}, {Thomas}, {Thomas}, {Troxel}, {Tucker}, {Vikram}, {Vivas},
  {Walker}, {Wechsler}, {Weller}, {Wester}, {Wolf}, {Wu}, {Yanny}, {Zenteno},
  {Zhang}, {Zuntz}, {DES Collaboration}, {Juneau}, {Fitzpatrick}, {Nikutta},
  {Nidever}, {Olsen}, {Scott}, \& {NOAO Data Lab}}]{2018ApJS..239...18A}
{Abbott}, T.~M.~C., {Abdalla}, F.~B., {Allam}, S., {et~al.} 2018, \apjs, 239,
  18, \dodoi{10.3847/1538-4365/aae9f0}

\bibitem[{{Abbott} {et~al.}(2021){Abbott}, {Adam{\'o}w}, {Aguena}, {Allam},
  {Amon}, {Annis}, {Avila}, {Bacon}, {Banerji}, {Bechtol}, {Becker},
  {Bernstein}, {Bertin}, {Bhargava}, {Bridle}, {Brooks}, {Burke}, {Carnero
  Rosell}, {Carrasco Kind}, {Carretero}, {Castander}, {Cawthon}, {Chang},
  {Choi}, {Conselice}, {Costanzi}, {Crocce}, {da Costa}, {Davis}, {De Vicente},
  {DeRose}, {Desai}, {Diehl}, {Dietrich}, {Drlica-Wagner}, {Eckert},
  {Elvin-Poole}, {Everett}, {Evrard}, {Ferrero}, {Fert{\'e}}, {Flaugher},
  {Fosalba}, {Friedel}, {Frieman}, {Garc{\'\i}a-Bellido}, {Gaztanaga},
  {Gelman}, {Gerdes}, {Giannantonio}, {Gill}, {Gruen}, {Gruendl}, {Gschwend},
  {Gutierrez}, {Hartley}, {Hinton}, {Hollowood}, {Honscheid}, {Huterer},
  {James}, {Jeltema}, {Johnson}, {Kent}, {Kron}, {Kuehn}, {Kuropatkin},
  {Lahav}, {Li}, {Lidman}, {Lin}, {MacCrann}, {Maia}, {Manning}, {Maloney},
  {March}, {Marshall}, {Martini}, {Melchior}, {Menanteau}, {Miquel}, {Morgan},
  {Myles}, {Neilsen}, {Ogando}, {Palmese}, {Paz-Chinch{\'o}n}, {Petravick},
  {Pieres}, {Plazas}, {Pond}, {Rodriguez-Monroy}, {Romer}, {Roodman}, {Rykoff},
  {Sako}, {Sanchez}, {Santiago}, {Scarpine}, {Serrano}, {Sevilla-Noarbe},
  {Smith}, {Smith}, {Soares-Santos}, {Suchyta}, {Swanson}, {Tarle}, {Thomas},
  {To}, {Tremblay}, {Troxel}, {Tucker}, {Turner}, {Varga}, {Walker},
  {Wechsler}, {Weller}, {Wester}, {Wilkinson}, {Yanny}, {Zhang}, {Nikutta},
  {Fitzpatrick}, {Jacques}, {Scott}, {Olsen}, {Huang}, {Herrera}, {Juneau},
  {Nidever}, {Weaver}, {Adean}, {Correia}, {de Freitas}, {Freitas},
  {Singulani}, {Vila-Verde}, \& {Linea Science Server}}]{2021ApJS..255...20A}
{Abbott}, T.~M.~C., {Adam{\'o}w}, M., {Aguena}, M., {et~al.} 2021, \apjs, 255,
  20, \dodoi{10.3847/1538-4365/ac00b3}

\bibitem[{{Astropy Collaboration} {et~al.}(2013){Astropy Collaboration},
  {Robitaille}, {Tollerud}, {Greenfield}, {Droettboom}, {Bray}, {Aldcroft},
  {Davis}, {Ginsburg}, {Price-Whelan}, {Kerzendorf}, {Conley}, {Crighton},
  {Barbary}, {Muna}, {Ferguson}, {Grollier}, {Parikh}, {Nair}, {Unther},
  {Deil}, {Woillez}, {Conseil}, {Kramer}, {Turner}, {Singer}, {Fox}, {Weaver},
  {Zabalza}, {Edwards}, {Azalee Bostroem}, {Burke}, {Casey}, {Crawford},
  {Dencheva}, {Ely}, {Jenness}, {Labrie}, {Lim}, {Pierfederici}, {Pontzen},
  {Ptak}, {Refsdal}, {Servillat}, \& {Streicher}}]{astropy}
{Astropy Collaboration}, {Robitaille}, T.~P., {Tollerud}, E.~J., {et~al.} 2013,
  \aap, 558, A33, \dodoi{10.1051/0004-6361/201322068}

\bibitem[{{Astropy Collaboration} {et~al.}(2018){Astropy Collaboration},
  {Price-Whelan}, {Sip{\H{o}}cz}, {G{\"u}nther}, {Lim}, {Crawford}, {Conseil},
  {Shupe}, {Craig}, {Dencheva}, {Ginsburg}, {VanderPlas}, {Bradley},
  {P{\'e}rez-Su{\'a}rez}, {de Val-Borro}, {Aldcroft}, {Cruz}, {Robitaille},
  {Tollerud}, {Ardelean}, {Babej}, {Bach}, {Bachetti}, {Bakanov}, {Bamford},
  {Barentsen}, {Barmby}, {Baumbach}, {Berry}, {Biscani}, {Boquien}, {Bostroem},
  {Bouma}, {Brammer}, {Bray}, {Breytenbach}, {Buddelmeijer}, {Burke},
  {Calderone}, {Cano Rodr{\'\i}guez}, {Cara}, {Cardoso}, {Cheedella}, {Copin},
  {Corrales}, {Crichton}, {D'Avella}, {Deil}, {Depagne}, {Dietrich}, {Donath},
  {Droettboom}, {Earl}, {Erben}, {Fabbro}, {Ferreira}, {Finethy}, {Fox},
  {Garrison}, {Gibbons}, {Goldstein}, {Gommers}, {Greco}, {Greenfield},
  {Groener}, {Grollier}, {Hagen}, {Hirst}, {Homeier}, {Horton}, {Hosseinzadeh},
  {Hu}, {Hunkeler}, {Ivezi{\'c}}, {Jain}, {Jenness}, {Kanarek}, {Kendrew},
  {Kern}, {Kerzendorf}, {Khvalko}, {King}, {Kirkby}, {Kulkarni}, {Kumar},
  {Lee}, {Lenz}, {Littlefair}, {Ma}, {Macleod}, {Mastropietro}, {McCully},
  {Montagnac}, {Morris}, {Mueller}, {Mumford}, {Muna}, {Murphy}, {Nelson},
  {Nguyen}, {Ninan}, {N{\"o}the}, {Ogaz}, {Oh}, {Parejko}, {Parley}, {Pascual},
  {Patil}, {Patil}, {Plunkett}, {Prochaska}, {Rastogi}, {Reddy Janga},
  {Sabater}, {Sakurikar}, {Seifert}, {Sherbert}, {Sherwood-Taylor}, {Shih},
  {Sick}, {Silbiger}, {Singanamalla}, {Singer}, {Sladen}, {Sooley},
  {Sornarajah}, {Streicher}, {Teuben}, {Thomas}, {Tremblay}, {Turner},
  {Terr{\'o}n}, {van Kerkwijk}, {de la Vega}, {Watkins}, {Weaver}, {Whitmore},
  {Woillez}, {Zabalza}, \& {Astropy Contributors}}]{astropy:2018}
{Astropy Collaboration}, {Price-Whelan}, A.~M., {Sip{\H{o}}cz}, B.~M., {et~al.}
  2018, \aj, 156, 123, \dodoi{10.3847/1538-3881/aabc4f}

\bibitem[{{Astropy Collaboration} {et~al.}(2022){Astropy Collaboration},
  {Price-Whelan}, {Lim}, {Earl}, {Starkman}, {Bradley}, {Shupe}, {Patil},
  {Corrales}, {Brasseur}, {N{\"o}the}, {Donath}, {Tollerud}, {Morris},
  {Ginsburg}, {Vaher}, {Weaver}, {Tocknell}, {Jamieson}, {van Kerkwijk},
  {Robitaille}, {Merry}, {Bachetti}, {G{\"u}nther}, {Aldcroft},
  {Alvarado-Montes}, {Archibald}, {B{\'o}di}, {Bapat}, {Barentsen},
  {Baz{\'a}n}, {Biswas}, {Boquien}, {Burke}, {Cara}, {Cara}, {Conroy},
  {Conseil}, {Craig}, {Cross}, {Cruz}, {D'Eugenio}, {Dencheva}, {Devillepoix},
  {Dietrich}, {Eigenbrot}, {Erben}, {Ferreira}, {Foreman-Mackey}, {Fox},
  {Freij}, {Garg}, {Geda}, {Glattly}, {Gondhalekar}, {Gordon}, {Grant},
  {Greenfield}, {Groener}, {Guest}, {Gurovich}, {Handberg}, {Hart},
  {Hatfield-Dodds}, {Homeier}, {Hosseinzadeh}, {Jenness}, {Jones}, {Joseph},
  {Kalmbach}, {Karamehmetoglu}, {Ka{\l}uszy{\'n}ski}, {Kelley}, {Kern},
  {Kerzendorf}, {Koch}, {Kulumani}, {Lee}, {Ly}, {Ma}, {MacBride}, {Maljaars},
  {Muna}, {Murphy}, {Norman}, {O'Steen}, {Oman}, {Pacifici}, {Pascual},
  {Pascual-Granado}, {Patil}, {Perren}, {Pickering}, {Rastogi}, {Roulston},
  {Ryan}, {Rykoff}, {Sabater}, {Sakurikar}, {Salgado}, {Sanghi}, {Saunders},
  {Savchenko}, {Schwardt}, {Seifert-Eckert}, {Shih}, {Jain}, {Shukla}, {Sick},
  {Simpson}, {Singanamalla}, {Singer}, {Singhal}, {Sinha}, {Sip{\H{o}}cz},
  {Spitler}, {Stansby}, {Streicher}, {{\v{S}}umak}, {Swinbank}, {Taranu},
  {Tewary}, {Tremblay}, {de Val-Borro}, {Van Kooten}, {Vasovi{\'c}}, {Verma},
  {de Miranda Cardoso}, {Williams}, {Wilson}, {Winkel}, {Wood-Vasey}, {Xue},
  {Yoachim}, {Zhang}, {Zonca}, \& {Astropy Project
  Contributors}}]{astropy:2022}
{Astropy Collaboration}, {Price-Whelan}, A.~M., {Lim}, P.~L., {et~al.} 2022,
  \apj, 935, 167, \dodoi{10.3847/1538-4357/ac7c74}

\bibitem[{{Banerjee} \& {Jog}(2011)}]{Banerjee2011}
{Banerjee}, A., \& {Jog}, C.~J. 2011, \apjl, 732, L8,
  \dodoi{10.1088/2041-8205/732/1/L8}

\bibitem[{{Belokurov} {et~al.}(2019){Belokurov}, {Deason}, {Erkal}, {Koposov},
  {Carballo-Bello}, {Smith}, {Jethwa}, \& {Navarrete}}]{Belokurov2019}
{Belokurov}, V., {Deason}, A.~J., {Erkal}, D., {et~al.} 2019, \mnras, 488, L47,
  \dodoi{10.1093/mnrasl/slz101}

\bibitem[{{Belokurov} \& {Koposov}(2016)}]{2016MNRAS.456..602B}
{Belokurov}, V., \& {Koposov}, S.~E. 2016, \mnras, 456, 602,
  \dodoi{10.1093/mnras/stv2688}

\bibitem[{Bovy {et~al.}(2012)Bovy, Rix, Liu, Hogg, Beers, \&
  Lee}]{bovy_spatial_2012}
Bovy, J., Rix, H.-W., Liu, C., {et~al.} 2012, The Astrophysical Journal, 753,
  148, \dodoi{10.1088/0004-637X/753/2/148}

\bibitem[{{Bowden} {et~al.}(2015){Bowden}, {Belokurov}, \&
  {Evans}}]{Bowden2015}
{Bowden}, A., {Belokurov}, V., \& {Evans}, N.~W. 2015, \mnras, 449, 1391,
  \dodoi{10.1093/mnras/stv285}

\bibitem[{{Bowden} {et~al.}(2016){Bowden}, {Evans}, \& {Williams}}]{Bowden2016}
{Bowden}, A., {Evans}, N.~W., \& {Williams}, A.~A. 2016, \mnras, 460, 329,
  \dodoi{10.1093/mnras/stw994}

\bibitem[{{Conroy} {et~al.}(2021){Conroy}, {Naidu}, {Garavito-Camargo},
  {Besla}, {Zaritsky}, {Bonaca}, \& {Johnson}}]{Conroy2021}
{Conroy}, C., {Naidu}, R.~P., {Garavito-Camargo}, N., {et~al.} 2021, \nat, 592,
  534, \dodoi{10.1038/s41586-021-03385-7}

\bibitem[{{Conroy} {et~al.}(2019){Conroy}, {Bonaca}, {Cargile}, {Johnson},
  {Caldwell}, {Naidu}, {Zaritsky}, {Fabricant}, {Moran}, {Rhee},
  {Szentgyorgyi}, {Berlind}, {Calkins}, {Kattner}, \&
  {Ly}}]{2019ApJ...883..107C}
{Conroy}, C., {Bonaca}, A., {Cargile}, P., {et~al.} 2019, \apj, 883, 107,
  \dodoi{10.3847/1538-4357/ab38b8}

\bibitem[{{Dark Energy Survey Collaboration} {et~al.}(2016){Dark Energy Survey
  Collaboration}, {Abbott}, {Abdalla}, {Aleksi{\'c}}, {Allam}, {Amara},
  {Bacon}, {Balbinot}, {Banerji}, {Bechtol}, {Benoit-L{\'e}vy}, {Bernstein},
  {Bertin}, {Blazek}, {Bonnett}, {Bridle}, {Brooks}, {Brunner}, {Buckley-Geer},
  {Burke}, {Caminha}, {Capozzi}, {Carlsen}, {Carnero-Rosell}, {Carollo},
  {Carrasco-Kind}, {Carretero}, {Castander}, {Clerkin}, {Collett}, {Conselice},
  {Crocce}, {Cunha}, {D'Andrea}, {da Costa}, {Davis}, {Desai}, {Diehl},
  {Dietrich}, {Dodelson}, {Doel}, {Drlica-Wagner}, {Estrada}, {Etherington},
  {Evrard}, {Fabbri}, {Finley}, {Flaugher}, {Foley}, {Fosalba}, {Frieman},
  {Garc{\'\i}a-Bellido}, {Gaztanaga}, {Gerdes}, {Giannantonio}, {Goldstein},
  {Gruen}, {Gruendl}, {Guarnieri}, {Gutierrez}, {Hartley}, {Honscheid}, {Jain},
  {James}, {Jeltema}, {Jouvel}, {Kessler}, {King}, {Kirk}, {Kron}, {Kuehn},
  {Kuropatkin}, {Lahav}, {Li}, {Lima}, {Lin}, {Maia}, {Makler}, {Manera},
  {Maraston}, {Marshall}, {Martini}, {McMahon}, {Melchior}, {Merson}, {Miller},
  {Miquel}, {Mohr}, {Morice-Atkinson}, {Naidoo}, {Neilsen}, {Nichol}, {Nord},
  {Ogando}, {Ostrovski}, {Palmese}, {Papadopoulos}, {Peiris}, {Peoples},
  {Percival}, {Plazas}, {Reed}, {Refregier}, {Romer}, {Roodman}, {Ross},
  {Rozo}, {Rykoff}, {Sadeh}, {Sako}, {S{\'a}nchez}, {Sanchez}, {Santiago},
  {Scarpine}, {Schubnell}, {Sevilla-Noarbe}, {Sheldon}, {Smith}, {Smith},
  {Soares-Santos}, {Sobreira}, {Soumagnac}, {Suchyta}, {Sullivan}, {Swanson},
  {Tarle}, {Thaler}, {Thomas}, {Thomas}, {Tucker}, {Vieira}, {Vikram},
  {Walker}, {Wechsler}, {Weller}, {Wester}, {Whiteway}, {Wilcox}, {Yanny},
  {Zhang}, \& {Zuntz}}]{2016MNRAS.460.1270D}
{Dark Energy Survey Collaboration}, {Abbott}, T., {Abdalla}, F.~B., {et~al.}
  2016, \mnras, 460, 1270, \dodoi{10.1093/mnras/stw641}

\bibitem[{{Das} {et~al.}(2023){Das}, {Ianjamasimanana}, {McGaugh}, {Schombert},
  \& {Dwarakanath}}]{Das2023}
{Das}, M., {Ianjamasimanana}, R., {McGaugh}, S.~S., {Schombert}, J., \&
  {Dwarakanath}, K.~S. 2023, \apjl, 946, L8, \dodoi{10.3847/2041-8213/acc10e}

\bibitem[{{Deason} {et~al.}(2011{\natexlab{a}}){Deason}, {Belokurov}, \&
  {Evans}}]{deason2011}
{Deason}, A.~J., {Belokurov}, V., \& {Evans}, N.~W. 2011{\natexlab{a}}, \mnras,
  416, 2903, \dodoi{10.1111/j.1365-2966.2011.19237.x}

\bibitem[{{Deason} {et~al.}(2011{\natexlab{b}}){Deason}, {Belokurov}, \&
  {Evans}}]{2011MNRAS.416.2903D}
---. 2011{\natexlab{b}}, \mnras, 416, 2903,
  \dodoi{10.1111/j.1365-2966.2011.19237.x}

\bibitem[{Deason {et~al.}(2011)Deason, Belokurov, \& Evans}]{deason_milky_2011}
Deason, A.~J., Belokurov, V., \& Evans, N.~W. 2011, Monthly Notices of the
  Royal Astronomical Society, 416, 2903,
  \dodoi{10.1111/j.1365-2966.2011.19237.x}

\bibitem[{{Deason} {et~al.}(2018){Deason}, {Belokurov}, \&
  {Koposov}}]{2018ApJ...852..118D}
{Deason}, A.~J., {Belokurov}, V., \& {Koposov}, S.~E. 2018, \apj, 852, 118,
  \dodoi{10.3847/1538-4357/aa9d19}

\bibitem[{{Deg} \& {Widrow}(2013)}]{Deg2013}
{Deg}, N., \& {Widrow}, L. 2013, \mnras, 428, 912, \dodoi{10.1093/mnras/sts089}

\bibitem[{{Dey} {et~al.}(2019){Dey}, {Schlegel}, {Lang}, {Blum}, {Burleigh},
  {Fan}, {Findlay}, {Finkbeiner}, {Herrera}, {Juneau}, {Landriau}, {Levi},
  {McGreer}, {Meisner}, {Myers}, {Moustakas}, {Nugent}, {Patej}, {Schlafly},
  {Walker}, {Valdes}, {Weaver}, {Y{\`e}che}, {Zou}, {Zhou}, {Abareshi},
  {Abbott}, {Abolfathi}, {Aguilera}, {Alam}, {Allen}, {Alvarez}, {Annis},
  {Ansarinejad}, {Aubert}, {Beechert}, {Bell}, {BenZvi}, {Beutler}, {Bielby},
  {Bolton}, {Brice{\~n}o}, {Buckley-Geer}, {Butler}, {Calamida}, {Carlberg},
  {Carter}, {Casas}, {Castander}, {Choi}, {Comparat}, {Cukanovaite}, {Delubac},
  {DeVries}, {Dey}, {Dhungana}, {Dickinson}, {Ding}, {Donaldson}, {Duan},
  {Duckworth}, {Eftekharzadeh}, {Eisenstein}, {Etourneau}, {Fagrelius},
  {Farihi}, {Fitzpatrick}, {Font-Ribera}, {Fulmer}, {G{\"a}nsicke},
  {Gaztanaga}, {George}, {Gerdes}, {Gontcho}, {Gorgoni}, {Green}, {Guy},
  {Harmer}, {Hernandez}, {Honscheid}, {Huang}, {James}, {Jannuzi}, {Jiang},
  {Joyce}, {Karcher}, {Karkar}, {Kehoe}, {Kneib}, {Kueter-Young}, {Lan},
  {Lauer}, {Le Guillou}, {Le Van Suu}, {Lee}, {Lesser}, {Perreault Levasseur},
  {Li}, {Mann}, {Marshall}, {Mart{\'\i}nez-V{\'a}zquez}, {Martini}, {du Mas des
  Bourboux}, {McManus}, {Meier}, {M{\'e}nard}, {Metcalfe},
  {Mu{\~n}oz-Guti{\'e}rrez}, {Najita}, {Napier}, {Narayan}, {Newman}, {Nie},
  {Nord}, {Norman}, {Olsen}, {Paat}, {Palanque-Delabrouille}, {Peng},
  {Poppett}, {Poremba}, {Prakash}, {Rabinowitz}, {Raichoor}, {Rezaie},
  {Robertson}, {Roe}, {Ross}, {Ross}, {Rudnick}, {Safonova}, {Saha},
  {S{\'a}nchez}, {Savary}, {Schweiker}, {Scott}, {Seo}, {Shan}, {Silva},
  {Slepian}, {Soto}, {Sprayberry}, {Staten}, {Stillman}, {Stupak}, {Summers},
  {Sien Tie}, {Tirado}, {Vargas-Maga{\~n}a}, {Vivas}, {Wechsler}, {Williams},
  {Yang}, {Yang}, {Yapici}, {Zaritsky}, {Zenteno}, {Zhang}, {Zhang}, {Zhou}, \&
  {Zhou}}]{2019AJ....157..168D}
{Dey}, A., {Schlegel}, D.~J., {Lang}, D., {et~al.} 2019, \aj, 157, 168,
  \dodoi{10.3847/1538-3881/ab089d}

\bibitem[{{Dillamore} {et~al.}(2022){Dillamore}, {Belokurov}, {Font}, \&
  {McCarthy}}]{Dillamore2022}
{Dillamore}, A.~M., {Belokurov}, V., {Font}, A.~S., \& {McCarthy}, I.~G. 2022,
  \mnras, 513, 1867, \dodoi{10.1093/mnras/stac1038}

\bibitem[{{Fellhauer} {et~al.}(2006){Fellhauer}, {Belokurov}, {Evans},
  {Wilkinson}, {Zucker}, {Gilmore}, {Irwin}, {Bramich}, {Vidrih}, {Wyse},
  {Beers}, \& {Brinkmann}}]{Fellhauer2006}
{Fellhauer}, M., {Belokurov}, V., {Evans}, N.~W., {et~al.} 2006, \apj, 651,
  167, \dodoi{10.1086/507128}

\bibitem[{{Feroz} {et~al.}(2009){Feroz}, {Hobson}, \&
  {Bridges}}]{2009MNRAS.398.1601F}
{Feroz}, F., {Hobson}, M.~P., \& {Bridges}, M. 2009, \mnras, 398, 1601,
  \dodoi{10.1111/j.1365-2966.2009.14548.x}

\bibitem[{{Flaugher} {et~al.}(2015){Flaugher}, {Diehl}, {Honscheid}, {Abbott},
  {Alvarez}, {Angstadt}, {Annis}, {Antonik}, {Ballester}, {Beaufore},
  {Bernstein}, {Bernstein}, {Bigelow}, {Bonati}, {Boprie}, {Brooks},
  {Buckley-Geer}, {Campa}, {Cardiel-Sas}, {Castander}, {Castilla}, {Cease},
  {Cela-Ruiz}, {Chappa}, {Chi}, {Cooper}, {da Costa}, {Dede}, {Derylo},
  {DePoy}, {de Vicente}, {Doel}, {Drlica-Wagner}, {Eiting}, {Elliott}, {Emes},
  {Estrada}, {Fausti Neto}, {Finley}, {Flores}, {Frieman}, {Gerdes},
  {Gladders}, {Gregory}, {Gutierrez}, {Hao}, {Holland}, {Holm}, {Huffman},
  {Jackson}, {James}, {Jonas}, {Karcher}, {Karliner}, {Kent}, {Kessler},
  {Kozlovsky}, {Kron}, {Kubik}, {Kuehn}, {Kuhlmann}, {Kuk}, {Lahav}, {Lathrop},
  {Lee}, {Levi}, {Lewis}, {Li}, {Mandrichenko}, {Marshall}, {Martinez},
  {Merritt}, {Miquel}, {Mu{\~n}oz}, {Neilsen}, {Nichol}, {Nord}, {Ogando},
  {Olsen}, {Palaio}, {Patton}, {Peoples}, {Plazas}, {Rauch}, {Reil}, {Rheault},
  {Roe}, {Rogers}, {Roodman}, {Sanchez}, {Scarpine}, {Schindler}, {Schmidt},
  {Schmitt}, {Schubnell}, {Schultz}, {Schurter}, {Scott}, {Serrano}, {Shaw},
  {Smith}, {Soares-Santos}, {Stefanik}, {Stuermer}, {Suchyta}, {Sypniewski},
  {Tarle}, {Thaler}, {Tighe}, {Tran}, {Tucker}, {Walker}, {Wang}, {Watson},
  {Weaverdyck}, {Wester}, {Woods}, {Yanny}, \& {DES Collaboration}}]{decam2015}
{Flaugher}, B., {Diehl}, H.~T., {Honscheid}, K., {et~al.} 2015, \aj, 150, 150,
  \dodoi{10.1088/0004-6256/150/5/150}

\bibitem[{Fukushima {et~al.}(2018)Fukushima, Chiba, Homma, Okamoto, Komiyama,
  Tanaka, Tanaka, Arimoto, \& Matsuno}]{fukushima_structure_2018}
Fukushima, T., Chiba, M., Homma, D., {et~al.} 2018, Publications of the
  Astronomical Society of Japan, Volume 70, Issue 4, id.69, 70, 69,
  \dodoi{10.1093/pasj/psy060}

\bibitem[{Fukushima {et~al.}(2019)Fukushima, Chiba, Tanaka, Hayashi, Homma,
  Okamoto, Komiyama, Tanaka, Arimoto, \& Matsuno}]{fukushima_stellar_2019}
Fukushima, T., Chiba, M., Tanaka, M., {et~al.} 2019, Publications of the
  Astronomical Society of Japan, 71, 72, \dodoi{10.1093/pasj/psz052}

\bibitem[{{Gaia Collaboration} {et~al.}(2023){Gaia Collaboration}, {Vallenari},
  {Brown}, {Prusti}, {de Bruijne}, {Arenou}, {Babusiaux}, {Biermann},
  {Creevey}, {Ducourant}, {Evans}, {Eyer}, {Guerra}, {Hutton}, {Jordi},
  {Klioner}, {Lammers}, {Lindegren}, {Luri}, {Mignard}, {Panem}, {Pourbaix},
  {Randich}, {Sartoretti}, {Soubiran}, {Tanga}, {Walton}, {Bailer-Jones},
  {Bastian}, {Drimmel}, {Jansen}, {Katz}, {Lattanzi}, {van Leeuwen}, {Bakker},
  {Cacciari}, {Casta{\~n}eda}, {De Angeli}, {Fabricius}, {Fouesneau},
  {Fr{\'e}mat}, {Galluccio}, {Guerrier}, {Heiter}, {Masana}, {Messineo},
  {Mowlavi}, {Nicolas}, {Nienartowicz}, {Pailler}, {Panuzzo}, {Riclet}, {Roux},
  {Seabroke}, {Sordo}, {Th{\'e}venin}, {Gracia-Abril}, {Portell}, {Teyssier},
  {Altmann}, {Andrae}, {Audard}, {Bellas-Velidis}, {Benson}, {Berthier},
  {Blomme}, {Burgess}, {Busonero}, {Busso}, {C{\'a}novas}, {Carry}, {Cellino},
  {Cheek}, {Clementini}, {Damerdji}, {Davidson}, {de Teodoro}, {Nu{\~n}ez
  Campos}, {Delchambre}, {Dell'Oro}, {Esquej}, {Fern{\'a}ndez-Hern{\'a}ndez},
  {Fraile}, {Garabato}, {Garc{\'\i}a-Lario}, {Gosset}, {Haigron}, {Halbwachs},
  {Hambly}, {Harrison}, {Hern{\'a}ndez}, {Hestroffer}, {Hodgkin}, {Holl},
  {Jan{\ss}en}, {Jevardat de Fombelle}, {Jordan}, {Krone-Martins}, {Lanzafame},
  {L{\"o}ffler}, {Marchal}, {Marrese}, {Moitinho}, {Muinonen}, {Osborne},
  {Pancino}, {Pauwels}, {Recio-Blanco}, {Reyl{\'e}}, {Riello}, {Rimoldini},
  {Roegiers}, {Rybizki}, {Sarro}, {Siopis}, {Smith}, {Sozzetti}, {Utrilla},
  {van Leeuwen}, {Abbas}, {{\'A}brah{\'a}m}, {Abreu Aramburu}, {Aerts},
  {Aguado}, {Ajaj}, {Aldea-Montero}, {Altavilla}, {{\'A}lvarez}, {Alves},
  {Anders}, {Anderson}, {Anglada Varela}, {Antoja}, {Baines}, {Baker},
  {Balaguer-N{\'u}{\~n}ez}, {Balbinot}, {Balog}, {Barache}, {Barbato},
  {Barros}, {Barstow}, {Bartolom{\'e}}, {Bassilana}, {Bauchet}, {Becciani},
  {Bellazzini}, {Berihuete}, {Bernet}, {Bertone}, {Bianchi}, {Binnenfeld},
  {Blanco-Cuaresma}, {Blazere}, {Boch}, {Bombrun}, {Bossini}, {Bouquillon},
  {Bragaglia}, {Bramante}, {Breedt}, {Bressan}, {Brouillet}, {Brugaletta},
  {Bucciarelli}, {Burlacu}, {Butkevich}, {Buzzi}, {Caffau}, {Cancelliere},
  {Cantat-Gaudin}, {Carballo}, {Carlucci}, {Carnerero}, {Carrasco},
  {Casamiquela}, {Castellani}, {Castro-Ginard}, {Chaoul}, {Charlot}, {Chemin},
  {Chiaramida}, {Chiavassa}, {Chornay}, {Comoretto}, {Contursi}, {Cooper},
  {Cornez}, {Cowell}, {Crifo}, {Cropper}, {Crosta}, {Crowley}, {Dafonte},
  {Dapergolas}, {David}, {David}, {de Laverny}, {De Luise}, {De March}, {De
  Ridder}, {de Souza}, {de Torres}, {del Peloso}, {del Pozo}, {Delbo},
  {Delgado}, {Delisle}, {Demouchy}, {Dharmawardena}, {Di Matteo}, {Diakite},
  {Diener}, {Distefano}, {Dolding}, {Edvardsson}, {Enke}, {Fabre}, {Fabrizio},
  {Faigler}, {Fedorets}, {Fernique}, {Fienga}, {Figueras}, {Fournier},
  {Fouron}, {Fragkoudi}, {Gai}, {Garcia-Gutierrez}, {Garcia-Reinaldos},
  {Garc{\'\i}a-Torres}, {Garofalo}, {Gavel}, {Gavras}, {Gerlach}, {Geyer},
  {Giacobbe}, {Gilmore}, {Girona}, {Giuffrida}, {Gomel}, {Gomez},
  {Gonz{\'a}lez-N{\'u}{\~n}ez}, {Gonz{\'a}lez-Santamar{\'\i}a},
  {Gonz{\'a}lez-Vidal}, {Granvik}, {Guillout}, {Guiraud},
  {Guti{\'e}rrez-S{\'a}nchez}, {Guy}, {Hatzidimitriou}, {Hauser}, {Haywood},
  {Helmer}, {Helmi}, {Sarmiento}, {Hidalgo}, {Hilger}, {H{\l}adczuk}, {Hobbs},
  {Holland}, {Huckle}, {Jardine}, {Jasniewicz}, {Jean-Antoine Piccolo},
  {Jim{\'e}nez-Arranz}, {Jorissen}, {Juaristi Campillo}, {Julbe}, {Karbevska},
  {Kervella}, {Khanna}, {Kontizas}, {Kordopatis}, {Korn}, {K{\'o}sp{\'a}l},
  {Kostrzewa-Rutkowska}, {Kruszy{\'n}ska}, {Kun}, {Laizeau}, {Lambert},
  {Lanza}, {Lasne}, {Le Campion}, {Lebreton}, {Lebzelter}, {Leccia}, {Leclerc},
  {Lecoeur-Taibi}, {Liao}, {Licata}, {Lindstr{\o}m}, {Lister}, {Livanou},
  {Lobel}, {Lorca}, {Loup}, {Madrero Pardo}, {Magdaleno Romeo}, {Managau},
  {Mann}, {Manteiga}, {Marchant}, {Marconi}, {Marcos}, {Marcos Santos},
  {Mar{\'\i}n Pina}, {Marinoni}, {Marocco}, {Marshall}, {Martin Polo},
  {Mart{\'\i}n-Fleitas}, {Marton}, {Mary}, {Masip}, {Massari},
  {Mastrobuono-Battisti}, {Mazeh}, {McMillan}, {Messina}, {Michalik}, {Millar},
  {Mints}, {Molina}, {Molinaro}, {Moln{\'a}r}, {Monari}, {Mongui{\'o}},
  {Montegriffo}, {Montero}, {Mor}, {Mora}, {Morbidelli}, {Morel}, {Morris},
  {Muraveva}, {Murphy}, {Musella}, {Nagy}, {Noval}, {Oca{\~n}a}, {Ogden},
  {Ordenovic}, {Osinde}, {Pagani}, {Pagano}, {Palaversa}, {Palicio},
  {Pallas-Quintela}, {Panahi}, {Payne-Wardenaar}, {Pe{\~n}alosa Esteller},
  {Penttil{\"a}}, {Pichon}, {Piersimoni}, {Pineau}, {Plachy}, {Plum}, {Poggio},
  {Pr{\v{s}}a}, {Pulone}, {Racero}, {Ragaini}, {Rainer}, {Raiteri}, {Rambaux},
  {Ramos}, {Ramos-Lerate}, {Re Fiorentin}, {Regibo}, {Richards}, {Rios Diaz},
  {Ripepi}, {Riva}, {Rix}, {Rixon}, {Robichon}, {Robin}, {Robin}, {Roelens},
  {Rogues}, {Rohrbasser}, {Romero-G{\'o}mez}, {Rowell}, {Royer}, {Ruz Mieres},
  {Rybicki}, {Sadowski}, {S{\'a}ez N{\'u}{\~n}ez}, {Sagrist{\`a} Sell{\'e}s},
  {Sahlmann}, {Salguero}, {Samaras}, {Sanchez Gimenez}, {Sanna},
  {Santove{\~n}a}, {Sarasso}, {Schultheis}, {Sciacca}, {Segol}, {Segovia},
  {S{\'e}gransan}, {Semeux}, {Shahaf}, {Siddiqui}, {Siebert}, {Siltala},
  {Silvelo}, {Slezak}, {Slezak}, {Smart}, {Snaith}, {Solano}, {Solitro},
  {Souami}, {Souchay}, {Spagna}, {Spina}, {Spoto}, {Steele},
  {Steidelm{\"u}ller}, {Stephenson}, {S{\"u}veges}, {Surdej}, {Szabados},
  {Szegedi-Elek}, {Taris}, {Taylor}, {Teixeira}, {Tolomei}, {Tonello}, {Torra},
  {Torra}, {Torralba Elipe}, {Trabucchi}, {Tsounis}, {Turon}, {Ulla}, {Unger},
  {Vaillant}, {van Dillen}, {van Reeven}, {Vanel}, {Vecchiato}, {Viala},
  {Vicente}, {Voutsinas}, {Weiler}, {Wevers}, {Wyrzykowski}, {Yoldas}, {Yvard},
  {Zhao}, {Zorec}, {Zucker}, \& {Zwitter}}]{2023A&A...674A...1G}
{Gaia Collaboration}, {Vallenari}, A., {Brown}, A.~G.~A., {et~al.} 2023, \aap,
  674, A1, \dodoi{10.1051/0004-6361/202243940}

\bibitem[{Gerhard(2012)}]{gerhard_dark_2012}
Gerhard, O. 2012, Proceedings of the International Astronomical Union, 8, 211,
  \dodoi{10.1017/S174392131300481X}

\bibitem[{{Gillessen} {et~al.}(2017){Gillessen}, {Plewa}, {Eisenhauer}, {Sari},
  {Waisberg}, {Habibi}, {Pfuhl}, {George}, {Dexter}, {von Fellenberg}, {Ott},
  \& {Genzel}}]{2017ApJ...837...30G}
{Gillessen}, S., {Plewa}, P.~M., {Eisenhauer}, F., {et~al.} 2017, \apj, 837,
  30, \dodoi{10.3847/1538-4357/aa5c41}

\bibitem[{Han {et~al.}(2022)Han, Conroy, Johnson, Speagle, Bonaca, Chandra,
  Naidu, Ting, Woody, \& Zaritsky}]{han_stellar_2022}
Han, J.~J., Conroy, C., Johnson, B.~D., {et~al.} 2022, The Astronomical
  Journal, 164, 249, \dodoi{10.3847/1538-3881/ac97e9}

\bibitem[{{Helmi}(2004)}]{Helmi2004}
{Helmi}, A. 2004, \mnras, 351, 643, \dodoi{10.1111/j.1365-2966.2004.07812.x}

\bibitem[{Helmi(2008)}]{helmi_stellar_2008}
Helmi, A. 2008, The Astronomy and Astrophysics Review, 15, 145,
  \dodoi{10.1007/s00159-008-0009-6}

\bibitem[{{Helmi}(2020)}]{helmi2020}
{Helmi}, A. 2020, \araa, 58, 205, \dodoi{10.1146/annurev-astro-032620-021917}

\bibitem[{Hernitschek {et~al.}(2018)Hernitschek, Cohen, Rix, Sesar, Martin,
  Magnier, Wainscoat, Kaiser, Tonry, Kudritzki, Hodapp, Chambers, Flewelling,
  \& Burgett}]{hernitschek_profile_2018}
Hernitschek, N., Cohen, J.~G., Rix, H.-W., {et~al.} 2018, The Astrophysical
  Journal, 859, 31, \dodoi{10.3847/1538-4357/aabfbb}

\bibitem[{Higson {et~al.}(2019)Higson, Handley, Hobson, \&
  Lasenby}]{higson_dynamic_2019}
Higson, E., Handley, W., Hobson, M., \& Lasenby, A. 2019, Statistics and
  Computing, 29, 891, \dodoi{10.1007/s11222-018-9844-0}

\bibitem[{{Hunter}(2007)}]{matplotlib}
{Hunter}, J.~D. 2007, Computing in Science and Engineering, 9, 90,
  \dodoi{10.1109/MCSE.2007.55}

\bibitem[{{Iorio} \& {Belokurov}(2019)}]{2019MNRAS.482.3868I}
{Iorio}, G., \& {Belokurov}, V. 2019, \mnras, 482, 3868,
  \dodoi{10.1093/mnras/sty2806}

\bibitem[{Iorio {et~al.}(2018)Iorio, Belokurov, Erkal, Koposov, Nipoti, \&
  Fraternali}]{iorio_first_2018}
Iorio, G., Belokurov, V., Erkal, D., {et~al.} 2018, Monthly Notices of the
  Royal Astronomical Society, 474, 2142, \dodoi{10.1093/mnras/stx2819}

\bibitem[{Koposov {et~al.}(2023)Koposov, Speagle, Barbary, Ashton, Bennett,
  Buchner, Scheffler, Cook, Talbot, Guillochon, Cubillos, Ramos, Johnson, Lang,
  Ilya, Dartiailh, Nitz, McCluskey, Archibald, Deil, Foreman-Mackey, Goldstein,
  Tollerud, Leja, Kirk, Pitkin, Sheehan, Cargile, Patel, \&
  Angus}]{koposov_joshspeagledynesty_2023}
Koposov, S., Speagle, J., Barbary, K., {et~al.} 2023, joshspeagle/dynesty:
  v2.1.2,  Zenodo, \dodoi{10.5281/zenodo.7995596}

\bibitem[{{Law} \& {Majewski}(2010)}]{Law2010}
{Law}, D.~R., \& {Majewski}, S.~R. 2010, \apj, 718, 1128,
  \dodoi{10.1088/0004-637X/718/2/1128}

\bibitem[{{Li} \& {S5 Collaboration}(2021)}]{S5DR1}
{Li}, T., \& {S5 Collaboration}. 2021, {Southern Stellar Stream Spectroscopic
  Survey: The First Public Data Release}, Data Release 1,  Zenodo,
  \dodoi{10.5281/zenodo.4695135}

\bibitem[{{Li} {et~al.}(2019){Li}, {Koposov}, {Zucker}, {Lewis}, {Kuehn},
  {Simpson}, {Ji}, {Shipp}, {Mao}, {Geha}, {Pace}, {Mackey}, {Allam}, {Tucker},
  {Da Costa}, {Erkal}, {Simon}, {Mould}, {Martell}, {Wan}, {De Silva},
  {Bechtol}, {Balbinot}, {Belokurov}, {Bland-Hawthorn}, {Casey}, {Cullinane},
  {Drlica-Wagner}, {Sharma}, {Vivas}, {Wechsler}, {Yanny}, \& {S5
  Collaboration}}]{2019MNRAS.490.3508L}
{Li}, T.~S., {Koposov}, S.~E., {Zucker}, D.~B., {et~al.} 2019, \mnras, 490,
  3508, \dodoi{10.1093/mnras/stz2731}

\bibitem[{{Majewski} {et~al.}(2003){Majewski}, {Skrutskie}, {Weinberg}, \&
  {Ostheimer}}]{2003ApJ...599.1082M}
{Majewski}, S.~R., {Skrutskie}, M.~F., {Weinberg}, M.~D., \& {Ostheimer}, J.~C.
  2003, \apj, 599, 1082, \dodoi{10.1086/379504}

\bibitem[{Medina {et~al.}(2018)Medina, Muñoz, Vivas, Carlin, Förster,
  Martínez, Galbany, González-Gaitán, Hamuy, Jaeger, Maureira, \&
  Martín}]{medina_discovery_2018}
Medina, G.~E., Muñoz, R.~R., Vivas, A.~K., {et~al.} 2018, The Astrophysical
  Journal, 855, 43, \dodoi{10.3847/1538-4357/aaad02}

\bibitem[{{Olling} \& {Merrifield}(2000)}]{Olling2000}
{Olling}, R.~P., \& {Merrifield}, M.~R. 2000, \mnras, 311, 361,
  \dodoi{10.1046/j.1365-8711.2000.03053.x}

\bibitem[{{Palau} \& {Miralda-Escud{\'e}}(2023)}]{Palau2023}
{Palau}, C.~G., \& {Miralda-Escud{\'e}}, J. 2023, \mnras, 524, 2124,
  \dodoi{10.1093/mnras/stad1930}

\bibitem[{{Pieres} {et~al.}(2020){Pieres}, {Girardi}, {Balbinot}, {Santiago},
  {da Costa}, {Carnero Rosell}, {Pace}, {Bechtol}, {Groenewegen},
  {Drlica-Wagner}, {Li}, {Maia}, {Ogando}, {dal Ponte}, {Diehl}, {Amara},
  {Avila}, {Bertin}, {Brooks}, {Burke}, {Carrasco Kind}, {Carretero}, {De
  Vicente}, {Desai}, {Eifler}, {Flaugher}, {Fosalba}, {Frieman},
  {Garc{\'\i}a-Bellido}, {Gaztanaga}, {Gerdes}, {Gruen}, {Gruendl}, {Gschwend},
  {Gutierrez}, {Hollowood}, {Honscheid}, {James}, {Kuehn}, {Kuropatkin},
  {Marshall}, {Miquel}, {Plazas}, {Sanchez}, {Serrano}, {Sevilla-Noarbe},
  {Sheldon}, {Smith}, {Soares-Santos}, {Sobreira}, {Suchyta}, {Swanson},
  {Tarle}, {Thomas}, {Vikram}, \& {Walker}}]{Pieres_2020}
{Pieres}, A., {Girardi}, L., {Balbinot}, E., {et~al.} 2020, \mnras, 497, 1547,
  \dodoi{10.1093/mnras/staa1980}

\bibitem[{{Prada} {et~al.}(2019){Prada}, {Forero-Romero}, {Grand}, {Pakmor}, \&
  {Springel}}]{Prada2019}
{Prada}, J., {Forero-Romero}, J.~E., {Grand}, R. J.~J., {Pakmor}, R., \&
  {Springel}, V. 2019, \mnras, 490, 4877, \dodoi{10.1093/mnras/stz2873}

\bibitem[{{Preston} {et~al.}(1991){Preston}, {Shectman}, \&
  {Beers}}]{Preston1991}
{Preston}, G.~W., {Shectman}, S.~A., \& {Beers}, T.~C. 1991, \apj, 375, 121,
  \dodoi{10.1086/170175}

\bibitem[{{Schlegel} {et~al.}(1998){Schlegel}, {Finkbeiner}, \&
  {Davis}}]{Schlegel:1998}
{Schlegel}, D.~J., {Finkbeiner}, D.~P., \& {Davis}, M. 1998, \apj, 500, 525,
  \dodoi{10.1086/305772}

\bibitem[{Searle \& Zinn(1978)}]{searle_composition_1978}
Searle, L., \& Zinn, R. 1978, The Astrophysical Journal, 225, 357,
  \dodoi{10.1086/156499}

\bibitem[{{Sesar} {et~al.}(2011){Sesar}, {Juri{\'c}}, \&
  {Ivezi{\'c}}}]{Sesar2011}
{Sesar}, B., {Juri{\'c}}, M., \& {Ivezi{\'c}}, {\v{Z}}. 2011, \apj, 731, 4,
  \dodoi{10.1088/0004-637X/731/1/4}

\bibitem[{{Sharma} {et~al.}(2023){Sharma}, {Bland-Hawthorn}, {Silk}, \&
  {Boehm}}]{sharma23}
{Sharma}, S., {Bland-Hawthorn}, J., {Silk}, J., \& {Boehm}, C. 2023, \mnras,
  521, 4074, \dodoi{10.1093/mnras/stad721}

\bibitem[{{Shipp} {et~al.}(2018){Shipp}, {Drlica-Wagner}, {Balbinot},
  {Ferguson}, {Erkal}, {Li}, {Bechtol}, {Belokurov}, {Buncher}, {Carollo},
  {Carrasco Kind}, {Kuehn}, {Marshall}, {Pace}, {Rykoff}, {Sevilla-Noarbe},
  {Sheldon}, {Strigari}, {Vivas}, {Yanny}, {Zenteno}, {Abbott}, {Abdalla},
  {Allam}, {Avila}, {Bertin}, {Brooks}, {Burke}, {Carretero}, {Castander},
  {Cawthon}, {Crocce}, {Cunha}, {D'Andrea}, {da Costa}, {Davis}, {De Vicente},
  {Desai}, {Diehl}, {Doel}, {Evrard}, {Flaugher}, {Fosalba}, {Frieman},
  {Garc{\'\i}a-Bellido}, {Gaztanaga}, {Gerdes}, {Gruen}, {Gruendl}, {Gschwend},
  {Gutierrez}, {Hartley}, {Honscheid}, {Hoyle}, {James}, {Johnson}, {Krause},
  {Kuropatkin}, {Lahav}, {Lin}, {Maia}, {March}, {Martini}, {Menanteau},
  {Miller}, {Miquel}, {Nichol}, {Plazas}, {Romer}, {Sako}, {Sanchez},
  {Santiago}, {Scarpine}, {Schindler}, {Schubnell}, {Smith}, {Smith},
  {Sobreira}, {Suchyta}, {Swanson}, {Tarle}, {Thomas}, {Tucker}, {Walker},
  {Wechsler}, \& {DES Collaboration}}]{2018ApJ...862..114S}
{Shipp}, N., {Drlica-Wagner}, A., {Balbinot}, E., {et~al.} 2018, \apj, 862,
  114, \dodoi{10.3847/1538-4357/aacdab}

\bibitem[{{Skilling}(2004)}]{2004AIPC..735..395S}
{Skilling}, J. 2004, in American Institute of Physics Conference Series, Vol.
  735, Bayesian Inference and Maximum Entropy Methods in Science and
  Engineering: 24th International Workshop on Bayesian Inference and Maximum
  Entropy Methods in Science and Engineering, ed. R.~{Fischer}, R.~{Preuss}, \&
  U.~V. {Toussaint}, 395--405, \dodoi{10.1063/1.1835238}

\bibitem[{Skilling(2006)}]{10.1214/06-BA127}
Skilling, J. 2006, Bayesian Analysis, 1, 833 , \dodoi{10.1214/06-BA127}

\bibitem[{{Smith} {et~al.}(2009){Smith}, {Evans}, \& {An}}]{Smith2009}
{Smith}, M.~C., {Evans}, N.~W., \& {An}, J.~H. 2009, \apj, 698, 1110,
  \dodoi{10.1088/0004-637X/698/2/1110}

\bibitem[{{Speagle}(2020)}]{2020MNRAS.493.3132S}
{Speagle}, J.~S. 2020, \mnras, 493, 3132, \dodoi{10.1093/mnras/staa278}

\bibitem[{Starkenburg {et~al.}(2019)Starkenburg, Youakim, Martin, Thomas,
  Aguado, Arentsen, Carlberg, González Hernández, Ibata, Longeard,
  McConnachie, Navarro, Sánchez-Janssen, \& Venn}]{starkenburg_pristine_2019}
Starkenburg, E., Youakim, K., Martin, N., {et~al.} 2019, Monthly Notices of the
  Royal Astronomical Society, 490, 5757, \dodoi{10.1093/mnras/stz2935}

\bibitem[{{Stringer} {et~al.}(2021){Stringer}, {Drlica-Wagner}, {Macri},
  {Mart{\'\i}nez-V{\'a}zquez}, {Vivas}, {Ferguson}, {Pace}, {Walker},
  {Neilsen}, {Tavangar}, {Wester}, {Abbott}, {Aguena}, {Allam}, {Bacon},
  {Bechtol}, {Bertin}, {Brooks}, {Burke}, {Carnero Rosell}, {Carrasco Kind},
  {Carretero}, {Costanzi}, {Crocce}, {da Costa}, {Pereira}, {De Vicente},
  {Desai}, {Diehl}, {Doel}, {Ferrero}, {Garc{\'\i}a-Bellido}, {Gaztanaga},
  {Gerdes}, {Gruen}, {Gruendl}, {Gschwend}, {Gutierrez}, {Hinton}, {Hollowood},
  {Honscheid}, {Hoyle}, {James}, {Kuehn}, {Kuropatkin}, {Li}, {Maia},
  {Marshall}, {Menanteau}, {Miquel}, {Morgan}, {Ogando}, {Palmese},
  {Paz-Chinch{\'o}n}, {Plazas}, {Roodman}, {Sanchez}, {Schubnell}, {Serrano},
  {Sevilla-Noarbe}, {Smith}, {Soares-Santos}, {Suchyta}, {Tarle}, {Thomas},
  {To}, {Varga}, {Wilkinson}, {Zhang}, \& {DES Collaboration}}]{Stringer_2021}
{Stringer}, K.~M., {Drlica-Wagner}, A., {Macri}, L., {et~al.} 2021, \apj, 911,
  109, \dodoi{10.3847/1538-4357/abe873}

\bibitem[{Thomas {et~al.}(2018)Thomas, McConnachie, Ibata, Côté, Martin,
  Starkenburg, Carlberg, Chapman, Fabbro, Famaey, Fantin, Gwyn,
  Hénault-Brunet, Malhan, Navarro, Robin, \& Scott}]{thomas_-type_2018}
Thomas, G.~F., McConnachie, A.~W., Ibata, R.~A., {et~al.} 2018, Monthly Notices
  of the Royal Astronomical Society, 481, 5223, \dodoi{10.1093/mnras/sty2604}

\bibitem[{{van der Walt} {et~al.}(2011){van der Walt}, {Colbert}, \&
  {Varoquaux}}]{numpy}
{van der Walt}, S., {Colbert}, S.~C., \& {Varoquaux}, G. 2011, Computing in
  Science and Engineering, 13, 22, \dodoi{10.1109/MCSE.2011.37}

\bibitem[{{Vera-Ciro} \& {Helmi}(2013)}]{Vera-Ciro2013}
{Vera-Ciro}, C., \& {Helmi}, A. 2013, \apjl, 773, L4,
  \dodoi{10.1088/2041-8205/773/1/L4}

\bibitem[{Virtanen {et~al.}(2020)Virtanen, Gommers, Oliphant, Haberland, Reddy,
  Cournapeau, Burovski, Peterson, Weckesser, Bright, {van der Walt}, Brett,
  Wilson, Millman, Mayorov, Nelson, Jones, Kern, Larson, Carey, Polat, Feng,
  Moore, {VanderPlas}, Laxalde, Perktold, Cimrman, Henriksen, Quintero, Harris,
  Archibald, Ribeiro, Pedregosa, {van Mulbregt}, \& {SciPy 1.0
  Contributors}}]{2020SciPy-NMeth}
Virtanen, P., Gommers, R., Oliphant, T.~E., {et~al.} 2020, Nature Methods, 17,
  261, \dodoi{10.1038/s41592-019-0686-2}

\bibitem[{Watkins {et~al.}(2009)Watkins, Evans, Belokurov, Smith, Hewett,
  Bramich, Gilmore, Irwin, Vidrih, Wyrzykowski, \&
  Zucker}]{watkins_substructure_2009}
Watkins, L.~L., Evans, N.~W., Belokurov, V., {et~al.} 2009, Monthly Notices of
  the Royal Astronomical Society, 398, 1757,
  \dodoi{10.1111/j.1365-2966.2009.15242.x}

\bibitem[{{Wenger} {et~al.}(2000){Wenger}, {Ochsenbein}, {Egret}, {Dubois},
  {Bonnarel}, {Borde}, {Genova}, {Jasniewicz}, {Lalo{\"e}}, {Lesteven}, \&
  {Monier}}]{Simbad}
{Wenger}, M., {Ochsenbein}, F., {Egret}, D., {et~al.} 2000, \aaps, 143, 9,
  \dodoi{10.1051/aas:2000332}

\bibitem[{Xue {et~al.}(2015)Xue, Rix, Ma, Morrison, Bovy, Sesar, \&
  Janesh}]{xue_radial_2015}
Xue, X.-X., Rix, H.-W., Ma, Z., {et~al.} 2015, The Astrophysical Journal, 809,
  144, \dodoi{10.1088/0004-637X/809/2/144}

\bibitem[{{Yanny} {et~al.}(2000){Yanny}, {Newberg}, {Kent},
  {Laurent-Muehleisen}, {Pier}, {Richards}, {Stoughton}, {Anderson}, {Annis},
  {Brinkmann}, {Chen}, {Csabai}, {Doi}, {Fukugita}, {Hennessy}, {Ivezi{\'c}},
  {Knapp}, {Lupton}, {Munn}, {Nash}, {Rockosi}, {Schneider}, {Smith}, \&
  {York}}]{yanny2000}
{Yanny}, B., {Newberg}, H.~J., {Kent}, S., {et~al.} 2000, \apj, 540, 825,
  \dodoi{10.1086/309386}

\end{thebibliography}
\bibliographystyle{aasjournal}



\appendix

\end{document}